\documentclass[12pt]{article}
\usepackage{latexsym}
\usepackage{graphicx}
\usepackage{color}
\usepackage[colorlinks,citecolor=magenta]{hyperref}
\usepackage{fullpage}
\usepackage[normalem,normalbf]{ulem}
\usepackage{natbib}
\usepackage[section]{placeins}
\setcitestyle{number,open={[},close={]}}
\usepackage{figlatex,etoolbox}
\usepackage{times}
\usepackage{setspace}
\usepackage{booktabs,caption}
\usepackage[flushleft]{threeparttable}
\usepackage{enumitem}
\newcommand{\subscript}[2]{$#1 _ #2$}

\newcommand{\specificthanks}[1]{\@fnsymbol{#1}}

\makeatletter
\patchcmd\@combinedblfloats{\box\@outputbox}{\unvbox\@outputbox}{}{%
  \errmessage{\noexpand\@combinedblfloats could not be patched}%
}%
\makeatother

\newcommand{\kishendel}{\bgroup\markoverwith{\textcolor{blue}{\rule[0.5ex]{2pt}{0.4pt}}}\ULon}

\usepackage[final]{changes}
\definechangesauthor[name={Subhashis Banerjee}, color=orange]{suban}
\definechangesauthor[name={Sourabh B Paul}, color=green]{sbp}
\definechangesauthor[name={Kishen Suraj P}, color=blue]{kishen}
\setremarkmarkup{(#2)}
\usepackage[disable]{todonotes}

\title{On monitoring development indicators using high resolution satellite images}

\author{Potnuru Kishen Suraj\thanks{Electrical Engineering, Indian Institute of Technology Delhi, New Delhi 110016} \and Ankesh Gupta\thanks{Computer Science and Engineering, Indian Institute of Technology Delhi, New Delhi 110016}$^{}$ \thanks{Both authors contributed equally to this work} \and Makkunda Sharma\footnotemark[2]$^{}$ \footnotemark[3] \and \href{http://hss.iitd.ac.in/faculty/sourabh-b-paul}{Sourabh Bikas Paul}\thanks{Humanities and Social Sciences, Indian Institute of Technology Delhi, New Delhi 110016} \and \href{http://www.cse.iitd.ac.in/~suban}{Subhashis Banerjee}\footnotemark[2]
}
\date{\today}

\begin{document}
\maketitle

\begin{abstract}
We develop a machine learning based tool for accurate prediction of socio-economic indicators from daytime satellite imagery. The diverse set of indicators are often not intuitively related to observable features in satellite images, and are not even always well correlated with each other. Our predictive tool is more accurate than using night light as a proxy, and can be used to predict missing data, smooth out noise in surveys, monitor development progress of a region, and flag potential anomalies. Finally, we use predicted variables to do robustness analysis of a regression study of high rate of stunting in India.
\end{abstract}

In a country like India, where there is a paucity of reliable and high frequency data,  evidence based design of policy interventions that are grounded on accurate estimates of economic and development indicators are difficult. Census data collection \citep{census} for the 1.2 billion population is cumbersome and expensive, and is carried out infrequently only about once in a decade. Census is also error prone and noisy due to the large variability in the data collection processes across the geography, and there is often no validation \citep{brown1971comparing, vemuri1994data, bose2008accuracy}. Smaller sample surveys \citep{nsso,nfhs4} tend to be more accurate, but they too are infrequent, and, in general, they do not comprehensively address all aspects of the economy. Availability of reliable estimates of economic and development indicators at high frequency may enable policy planning in several ways - for example in block level development planning, addressing local issues related to agriculture, health, sanitation, education, employment, connectivity, resource management etc. Traditionally, extrapolated data estimates or inferential exercises are sometimes used for such purposes \citep{barro2013new}.  In this paper we explore the prospect of using predicted data as a viable alternative to using real data \citep{WPS7841}.

 We investigate the possibility of accurate regression of economic indicators, at the village level, on high resolution daytime satellite imagery, which can be acquired at low cost, accurately and frequently.  Our work is motivated by \citep{Jean790} who introduced deep CNN based prediction of development indicators from satellite images. We extend their work by not only  covering a ground area of 7 {\it Km}$^2$ on the average - which is at a resolution higher by at least one order of magnitude - in a much more diverse geography, but also by demonstrating accurate prediction of a much wider set of indicators including assets, infrastructure, demographics, education and health.

We carry out village level analysis in six Indian states - Punjab, Haryana, Uttar Pradesh, Bihar, Jharkhand and West Bengal. We ignore the urban areas. The problem is challenging not only because of large variability in language, culture and living styles across these states, but even the geographical diversity is significant. The topography varies from the agricultural plains in Punjab and Haryana to the Indo-Gangetic plains in the middle to the dense forests in parts of Jharkhand to the Chota Nagpur plateau to the mountains in the north and the  Sunder Bans delta in West Bengal. The differences, both in the socio-economic patterns and the visual characteristics, make the regression problem interesting. Recent advances in deep convolutional neural networks (CNNs) \citep{NIPS2012_4824,Goodfellow-et-al-2016} and GPU computing make the nonlinear regression problem addressable. We investigate both the accuracy of regression and what makes it possible.

Specifically, the main contributions of our paper are as follows:
\begin{enumerate}
\item We train a eight layer deep convolutional neural network based model (VGG CNN-S) \citep{Chatfield14}  for direct regression of census asset indicators on high resolution daytime satellite images of a village. We obtain superior  regression scores\footnote{All $R^2$ scores reported in this paper, except for the OLS study in Section \ref{sec:stunting}, are for out-of-sample data.} when compared to transfer learning \citep{5288526} from regression of night light data, a popular proxy economic indicator suggested in the literature \citep{Jean790}.
\item The regression model learnt from the data of 2,18,000 villages across six states smoothes out the error in the census data. This not only gives us accurate prediction of the census indicators, but we also obtain a census validation tool as a by-product.
\item We show that the asset prediction model can be effectively used for transfer learning \citep{5288526} of other socio-economic indicators for literacy, education, health and demographics.
\item Though the prediction model is trained with cross-sectional data, it can still facilitate monitoring  progress of development  of a region over time.
\item We show that spatial discontinuities (or sharp gradients) in the regression output over a geographical region can potentially indicate dissimilar outcomes of different policy interventions, and can serve as effective alerts for further investigation.
\item Finally, we demonstrate the utility of our tool by carrying out a case study of regression analysis using predicted variables to understand the determinants of the high rate of stunting in Indian children. We show that the predicted variables are helpful to avoid omitted variable biases.
\end{enumerate}

The rest of the paper is organized as follows. In Section \ref{sec:ML} we briefly discuss the considerations for using predictive machine learning models for econometrics. In Section \ref{sec:prior} we discuss some prior work that use satellite images for econometric analysis.  In Section \ref{sec:regression} we discuss our CNN based regression framework  for  transfer learning from night light data and direct regression of the asset indicators. In Section \ref{sec:transfer} we discuss transfer learning of other socio-economic parameters. In Sections \ref{sec:temporal} and \ref{sec:rd} we discuss monitoring development over time and spatial discontinuities in regression output respectively. In Section \ref{sec:stunting} we present our case study of regression analysis to understand the determinants of stunting.  We briefly describe our data sources in the appendix in Section \ref{sec:data}.

\section{Machine learning, prediction and causation}
\label{sec:ML}

Our machine learning based regression model is predictive and it does not directly support any causal inference. However, as has been pointed out in \citep{10.1257/aer.p20151023}, deciding on effective policy interventions require both prediction and causation. 
For example, to maximize an appropriate payoff function of a policy intervention targeted towards literacy, we may need to do both - accurately predict the literacy rate in a village, and evaluate how a policy intervention may affect literacy. Whereas, when the outcome is an outbreak of a disease like cholera in a village, where the causal relations and the policy implications are already reasonably well understood, a payoff maximization crucially depends on accurate and timely prediction of a possible outbreak. Accurate predictions can not only have large policy impacts but can also provide crucial theoretical and economic insights \citep{10.1257/aer.p20151023}, can offer the possibility of using predicted variables when real measurements are not available \citep{WPS7841}, and can be used to guard against omitted variable biases in econometric analysis \citep{barreto2006introductory}. Machine learning techniques can also be used in econometrics  to accurately predict an endogenous variable at the first stage of a linear instrumental regression model \citep{mullainathan2017machine}. \cite{athey2017state} review some innovative applications of machine learning methods in causal econometric inferences.

The value of accurate prediction has  been underemphasized in conventional econometrics. There are two main reasons. First, standard empirical methods are often based on small or infrequent sample surveys which provide inadequate data for accurate prediction estimates, and, consequently, the extrapolations are usually noisy. Building highly accurate predictive models will remain difficult till progress in digitzation can facilitate online, real-time gathering of high volume transactional socio-economic and health data. Second, standard linear regression techniques like OLS, which minimize the in-sample error, are not best suited for prediction problems because of their emphasis on reducing bias  at a cost of prediction accuracy \citep{10.1257/aer.p20151023}. They typically under-fit the data. In contrast, modern machine learning techniques like deep CNNs \citep{NIPS2012_4824,Chatfield14} explicitly address the bias-variance tradeoff  by a) using a nonlinear representational network which can model highly complex functions  b)  focusing on minimizing out of sample prediction error using $n$-fold cross validation  during the training phase  c)  incorporating regularization functions in the training optimization model that minimizes variance and d)  selectively using techniques like dropout to avoid over-fitting \citep{hastie_09_elements-of.statistical-learning, Goodfellow-et-al-2016}.  Hence, they are tuned for accurate prediction though the interpretability of the trained network model is low. 

Despite these advantages machine learning models have found limited use in econometrics mainly because of their large data requirement. 
However, CNN based prediction using satellite images provide more opportunities.  For example, the deep CNN model  VGG CNN-S  \citep{Chatfield14} that we use in this work has approximately 138 million parameters and has been trained to recognize 1000 image categories using a training set of 1.3 million labelled images, a validation set of 50,000 images and a test set of 100,000 images. Such a convolutional network, pre-trained with  large volumes of data, has already learned complex image representations in terms of highly discriminatory image features, which enables us to fine tune the model for regression of economic indicators using  a relatively small number of images.

Another significant difference between conventional econometric regression and deep CNN based regression manifests in the method of choosing the explanatory variables to predict the outcome. In both conventional econometrics and traditional machine learning the explanatory variables, or the features used for regression, are usually hand-crafted, and are measured from the data using well specified procedures. This makes the explanatory variables clearly identifiable, and the importance of each explanatory variable in predicting the outcome can be evaluated. In  contrast, a deep CNN automatically learns the best discriminatory features for the task at hand. The feature representations are distributed over the weights of the CNN making the features unidentifiable, and, consequently, any subsequent causal reasoning is difficult.  However, there have been some recent progress in explanatory understanding of knowledge representation in CNNs \citep{DBLP:journals/corr/SimonyanVZ13,Mahendran:2016:VDC:2995936.2995953,DBLP:journals/corr/abs-1708-01785,DBLP:journals/corr/WeiZTF15,DBLP:journals/corr/ZeilerF13} which may eventually lead to better interpretability of CNN models and facilitate causal analysis.

\section{Prior work}
\label{sec:prior}

Most of the prior work using satellite images use night light as a proxy for development. See \citep{su5124988} for a review. Night lights have been used to create a global grid of economic activity  \citep{openG2010};  to show that during 1992-2009 the centre of gravity of economic activity has shifted towards south and east as economies in India, China and Southeast Asia have `lit' up \citep{Cauwels};  to create a world poverty map using World Development Indicators (WDI) 2006 national level estimates for calibration \citep{ELVIDGE20091652}; as a measure of poverty to enable health interventions \citep{Noor2008}; to estimate the informal economy using night light images \citep{Ghosh_2009}; to develop an alternative measure of distribution of wealth \citep{sg-7-23-2012}; and even for estimating an information and technology development index \citep{ghosh-2010}. Most of the above methods use linear regression models using night lights \citep{su5124988}. In some recent India related studies, \cite{bhandari} show that night light is a valid proxy for economic activity in India, \cite{gst-epw} use night light images to show that inequality in India has been growing both within and across states, and \cite{asher-novosad} use night light images and regression discontinuities to show that ruling political dispensations tend to favour regions that represent their members. 

Daytime satellite images, of resolutions varying from $30m \times 30m$ to $0.5m \times 0.5m$ have also been used for econometric analysis. \cite{doi:10.1093/qje/qjs034} use satellite images to measure deforestation in Indonesia; \cite{10.2307/20648925} estimates the impact of air pollution (particulate matter) resulting from Indonesia's devastating late-1997 forest fire on infant and fetal mortality;  \cite{kenya-roof} use high resolution  $0.5m \times 0.5m$ satellite images to measure shiny roof as a proxy for dwelling investments in a Nairobi slum and \cite{RePEc:ucp:jpolec:doi:10.1086/684719} estimate economic impacts of climate change in agriculture. All the above methods compute hand-crafted features from satellite images and most use linear regression. See \citep{10.1257/jep.30.4.171} for an excellent survey and a primer on remote sensing for economists. See \citep{RePEc:eee:regchp:5-115, 10.1257/jep.28.2.3} for overviews of econometrics issues associated with spatial data and large datasets.

Use of machine learning techniques on satellite images for econometric analysis is relatively new. \cite{Albert2017} use deep CNN and satellite imagery to identify land use patterns. The economic survey of 2016-2017 of the finance ministry of government of India use satellite images to calculate built-up area and estimate potential property tax collection \citep{econsurvey1}. They use Principal Component Analysis (PCA) in conjugation with an ensemble model based on  Gradient Boosting Model (GBM) algorithm \citep{GBM} and multinomial classification regression (softmax) \citep{ufldl}. They extend this work in the second volume of the economic survey \citep{econsurvey2} to show that India may be more urbanized than previously thought.

Our work is motivated by \citep{Jean790} who use machine learning on daytime satellite images to predict poverty. They use a transfer learning approach \citep{5288526} to first fine tune VGG CNN-F \citep{Chatfield14}, a deep CNN model pre-trained to recognize 1000 image categories, to predict the average night light corresponding to a region from high resolution daytime satellite images. Night light is used here as a noisy but easily obtainable proxy for poverty. This fine tuning builds a deep CNN model that learns to predict economic activity from daytime images. The output vectors of the last fully connected layer (fc7) are then used as input features for regression of asset and consumption  using data obtained through sample surveys. The low volume sample survey data turns out to be adequate for the last regression step. 

We first reproduce the results of \citep{Jean790} to fine tune VGG CNN-S \citep{Chatfield14} to predict night light corresponding to 2,18,000 villages in the six Indian states. We then follow the same approach of fine tuning VGG CNN-S  to directly predict an asset vector obtained out of the Census 2011 \citep{census} data for these villages. We find that the latter approach works better not only for prediction of assets but also for transfer learning \citep{5288526} of other economic indicators.

\section{Predicting asset indicators using deep CNN}
\label{sec:regression}
We create an asset vector corresponding to each village from the {\it Houselisting and Housing} data (Table HH-14) of Census 2011 \citep{census} and obtain the corresponding images  from Google static maps using the \cite{googleapi}\footnote{The Google static maps only has recent imagery corresponding to 2017.}. We obtain the night light data from the Defense Meteorological Satellite Program's Operational Linescan System (DMSP-OLS) \citep{night}. Please see Section \ref{sec:data} in the appendix for details.

We use the pre-trained VGG CNN-S convolutional neural network  \citep{Chatfield14},  and modify it for our regression tasks as described below.   We convert the standard  input image size  of $224 \times 224$ of the model to $400 \times 400$. Consequently, only the pre-learnt weights of the convolutional layers could be reused in the new model, and the three fully connected layers needed to be trained afresh for which weights were initialized with a zero mean Gaussian distribution. Changing the underlying task of the deep CNN from classification to regression necessitated change of the hyper-parameters of the model. The weight decay (regularization parameter) was changed from $0.0005$ to $0.005$. Caffe \citep{DBLP:journals/corr/JiaSDKLGGD14} does not  support regression by default, so custom layers for data and $R^2$ had to be created. The weights were learnt using Stochastic Gradient Descent (SGD) \citep{ufldl}  with a batch size of 32 on a K40 NVidia GPU. The SGD was performed using a step learning policy with a learning rate of $10^{-6}$, $\gamma$ of $0.2$ and momentum of $0.8$.

\subsection{Regression of night light data} 
\label{sec:regnightlight}
We first train the deep CNN model to predict night light values from daytime satellite imagery.  We  identify the set of relevant night light cells and collect  the  daytime satellite images  corresponding to the centre of these cells. We limit the search space by using ESRI shape files of India available from GADM \citep{gadm}. Out of the approximately 4 million night cells corresponding to India, nearly 50\% have zero value resulting in a skewed night light histogram. We reduce the skew (third order moment) from 3.62 to 0.4 by under-sampling to obtain a final dataset of 2,19,000 night cell data points. We perform the regression   with a train-test split of 8:2.

We experiment with two different sizes of daytime images of $400 \times 400$ and $640 \times 640$, both at a zoom level of 16. At this zoom level they cover approximately 1 {\it Km}$^2$ and 2.5 {\it Km}$^2$ respectively, and they both subsume a night light cell which covers less than 1 {\it Km}$^2$.  We map the daytime images to the input size of $400 \times 400$  required by the modified VGG CNN-S, and build a custom data layer at the input for on-the-fly data augmentations \citep{Goodfellow-et-al-2016} like flip, vertical flip and image rotations etc.

We obtain regression $R^2$ scores of 0.69 and 0.79 for 1 {\it Km}$^2$ and 2.5 {\it Km}$^2$ respectively on the test set. The superior regression accuracy on increasing the ground area can perhaps be explained by the fact that light has a spread; and night light at a particular cell is influenced by the habitation in the surroundings cells as well, and that the neighbourhood context is important for predicting the value at the current cell.

\subsection{Transfer learning of asset vectors from the night light model}
Can the CNN model trained to predict night light from images be used to predict other economic indicators?
We  use the night light CNN model to predict a 16 dimensional vector representing the household level assets in a village. We create the 16 dimensional asset vector from census data as described in Table \ref{table:assets} (Section ~\ref{sec:data}) in the appendix.

\begin{figure}[h!]
	\centering
    \includegraphics[width=0.8\textwidth]{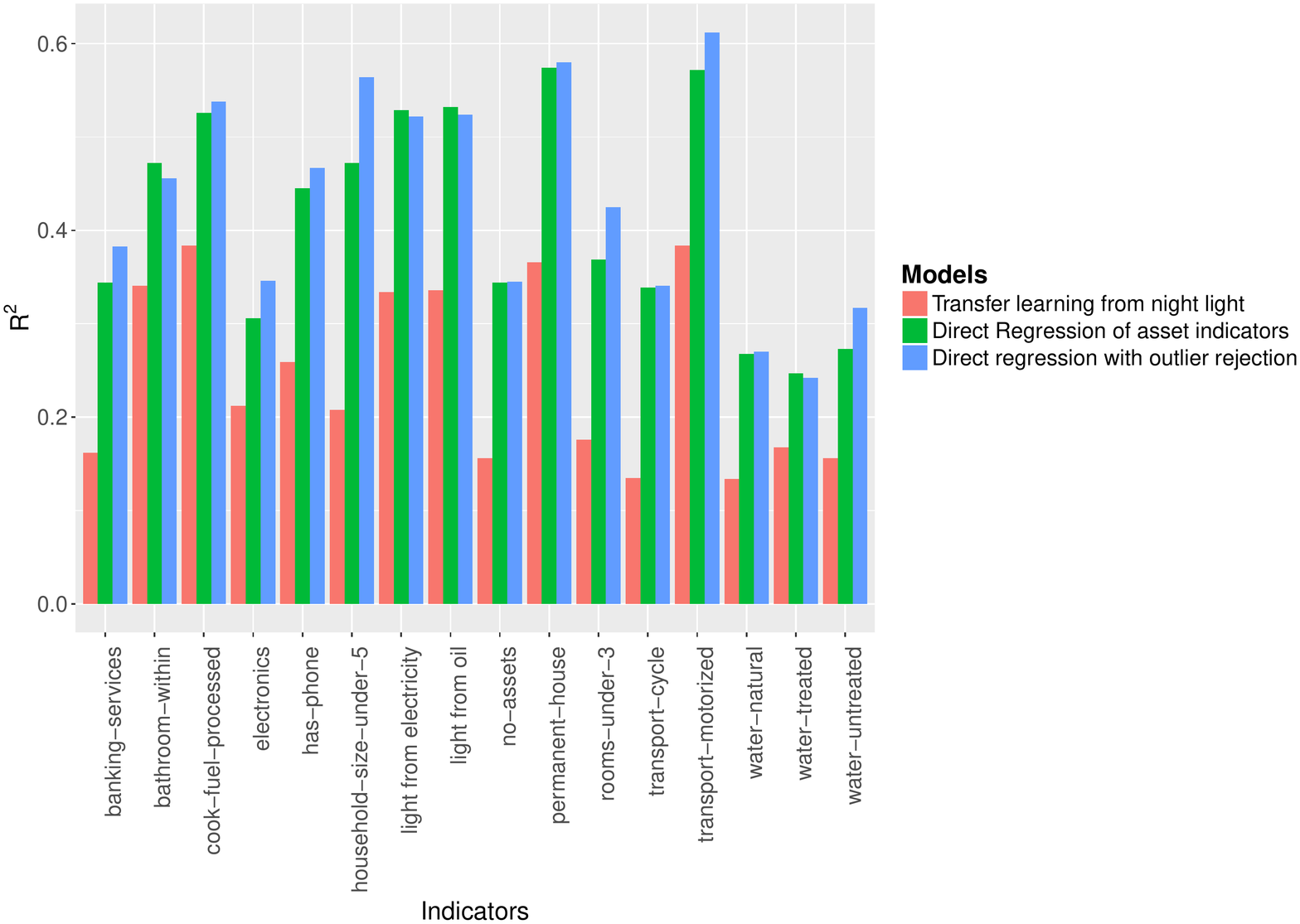}
    \caption{Comparison of out-of-sample $R^2$ scores for prediction of each indicator of the asset vector for the three methods.}
    \label{fig:regression}
\end{figure}

\begin{figure}[h!]
	\centering
    \includegraphics[width=0.8\textwidth]{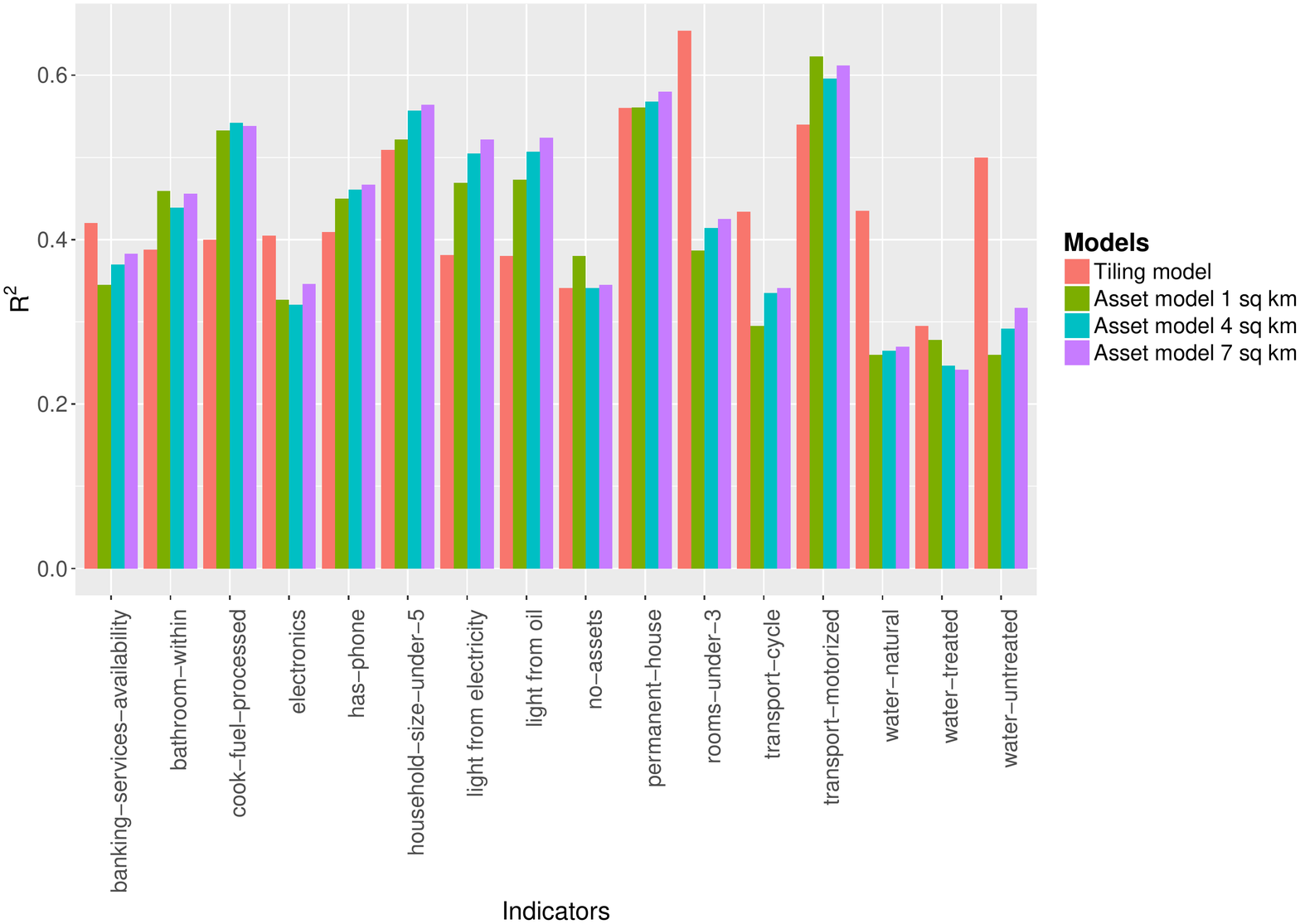}
    \caption{Comparison of out-of-sample $R^2$ scores for prediction of each indicator of the asset vector by different methods and parameters.}
    \label{fig:multiregression}
\end{figure}

The output of the night light model at the last hidden layer serves as a feature vector which provides a 4096 dimensional representation  of the village's economic prosperity, captured by the night light as a proxy. We use this for transfer learning \citep{5288526} of the asset vectors. We use the $640 \times 640$ daytime images at a zoom level  of 16 as input, and design a neural network with a single fully connected layer with rectified linear activation \citep{Goodfellow-et-al-2016} to do a regression of the asset vectors on the 4096 dimensional representation. A typical Indian village has an area between 5 to 7 $Km^2$, and we create the 4096 dimensional representation by taking the average of the output corresponding to four overlapping input images, each of size $640 \times 640$ covering roughly 2.5 $Km^2$, to cover approximately 7 $ Km^2$ centered at the village. The prediction accuracy of the transfer learning, though reasonable (Figure \ref{fig:regression}), is lower than what has been reported in \citep{Jean790}.  Since night light as a proxy for economic development has been reported to work well in India \citep{chen-nordhaus, bhandari, gst-epw, su5124988},  and we can predict night light quite accurately, the low regression scores of transfer learning can perhaps be ascribed to the noise in the census data.

\subsection{Direct regression of the asset vector}
\label{sec:regdirect}

\begin{figure}[h!]
\begin{tabular}{cc}
\includegraphics[width=3.0in]{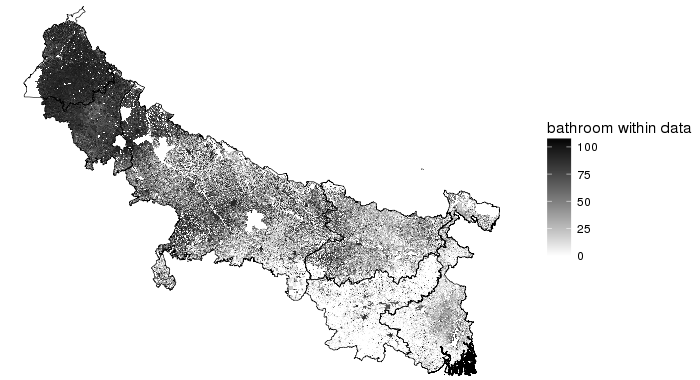} & \includegraphics[width=3.0in]{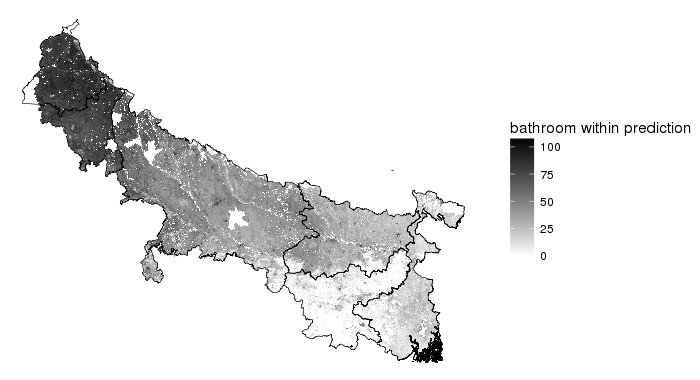}  \\
\includegraphics[width=3.0in]{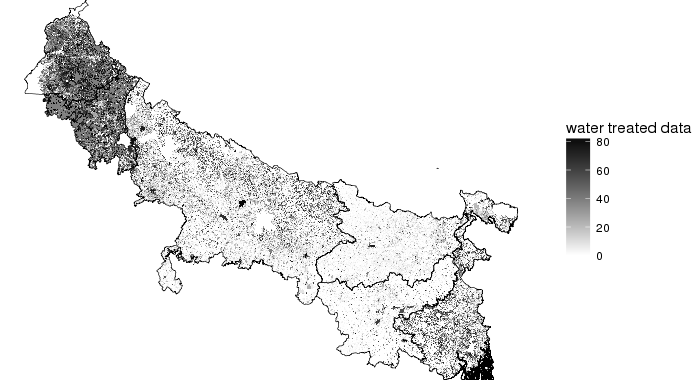} & \includegraphics[width=3.0in]{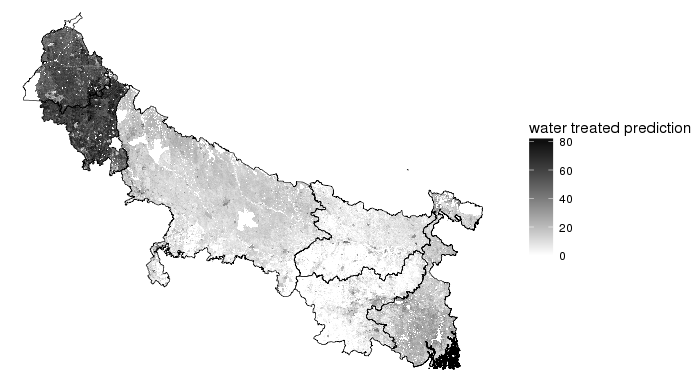}  \\
\includegraphics[width=3.0in]{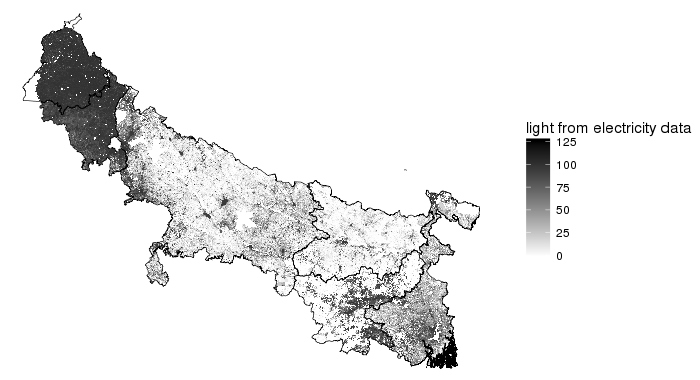} & \includegraphics[width=3.0in]{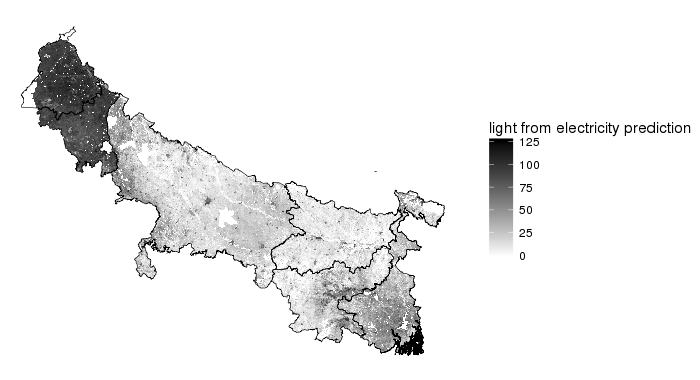}  \\
\includegraphics[width=3.0in]{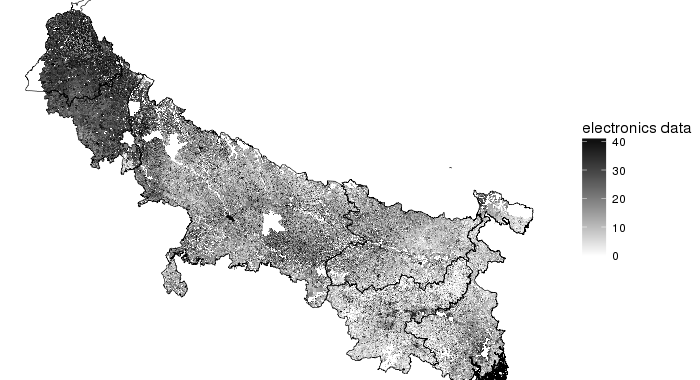} & \includegraphics[width=3.0in]{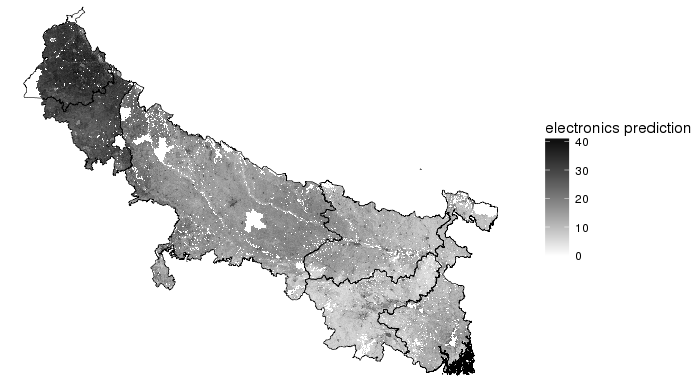}  \\
(a) Census data & (b) Prediction  \\
\end{tabular}
\caption{Choropleths for census data and model prediction for the `bathroom-within', `water-treated',  `light from electricity' and `electronics'  indicators. The values are percentages. The predicted output is sometimes greater than 100\%. See \href{http://web.iitd.ac.in/~suban/satellite/asset-model/}{http://web.iitd.ac.in/$\sim$suban/satellite/asset-model/} for the choropleths  for the other indicators and of the $R^2$ scores. The choropleths for the $R^2$ scores show that there is no significant regional bias in the prediction. The gaps in the choropleths are due to urban regions which we do not consider in our analysis.} 
\label{fig:reg}
\end{figure}

\begin{figure}[h!]
\begin{tabular}{cc}
\includegraphics[width=3.0in]{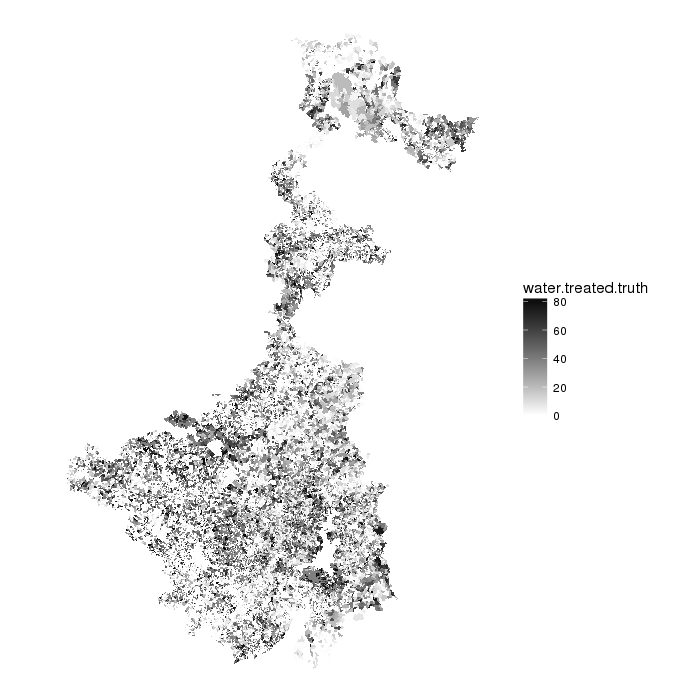} & \includegraphics[width=3.0in]{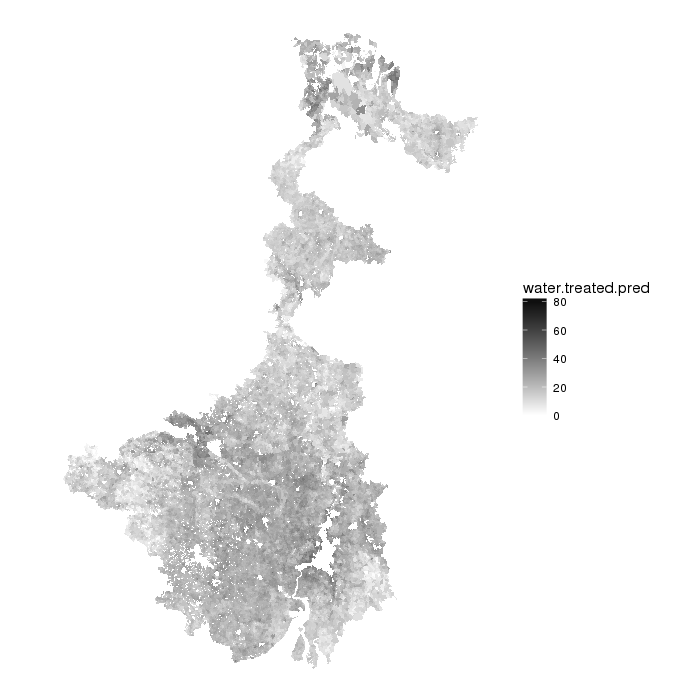}  \\
(a) Census data & (b) Prediction  \\
\end{tabular}
\caption{Census data and model prediction for the `water-treated' indicator for West Bengal. The values are percentages.}
\label{fig:regwb}
\end{figure}
Availability of a  large dataset with  census data corresponding to 2,18,000 villages from six states allows us to train a deep CNN for direct regression of the asset vectors on daytime images.  We modify the last fully connected layer to have a 16 dimensional output corresponding to the asset vectors described in Section \ref{sec:data} in appendix.

We define the objective function for training the regression model as
\[
C(\theta) = \frac{1}{2M} \sum_{i=1}^M \sum_{j=1}^{16} (f_i^j - y_i^j)^2 + \frac{d}{2} \sum_l d_l  \sum {w^l}^2
\]
where $\theta$ is the vector of all parameters (weights) in the model; the first term is the Euclidean loss where $f_i^j$ is the prediction value of the $j^{th}$ indicator for the $i^{th}$ village and $y_i^j$ is the value computed from the census data;  the second is a $L2$ regularization term  where $d$ is a {\it weight decay}, $d_l$ is the {\it decay multiplier} for the layer $l$ and the last sum is over all weights ${w^l}$ in the layer. We use $d = 0.005$ and $d_l = $1.  We initialize the weights in the last three fully connected layers  using a Gaussian distribution with zero mean and standard deviation of 0.005. The remaining layers are initialized  with the pre-learnt values of VGG CNN-S. During training  all weights of all layers are updated during the optimization process.  We use Caffe \citep{DBLP:journals/corr/JiaSDKLGGD14} for specification and training of the network.

The input daytime satellite imagery was collected via Google static maps \citep{googleapi} at a zoom level of 15 and a size of $640 \times 640$ which corresponds to a ground area of 7 {\it km}$^2$, which covers 95\% of all villages. The training and test set split used  was 8:2.

We also experimented with varying the input image sizes to cover 1, 4 and 7 {\it Km}$^2$ of the ground area; and with tiling each village with 1  {\it Km}$^2$ tiles and training each image tile with the aggregate village census data.  In Figure \ref{fig:multiregression} we show the per-indicator $R^2$ scores for various choices. The tiling model and the choice of 7 {\it km}$^2$ were comparable and gave better $R^2$ scores for regression. We choose the simpler 7 {\it km}$^2$ input image size for subsequent experiments.

Since the census data is noisy, we train the network both with and without outlier rejection (described in Section \ref{sec:data} in appendix). As a result of outlier rejection 17,000 villages are removed from the dataset resulting in a reduced dataset of 2,01,000 villages. The mean Euclidean loss for the 17,000 outlier points was 14,800 which is significantly higher than the mean overall Euclidean loss of 5,188 corresponding to the final model. The mean overall Euclidean loss without outlier rejection was 5,944. In Figure \ref{fig:regression} we compare the direct regression $R^2$ scores, with and without outlier rejection,  with that of transfer learning from night light regression.

As can be noted from Figures \ref{fig:regression} and   \ref{fig:multiregression}, the prediction accuracy of direct regression is superior to what is obtained from transfer learning from training a night light model. 

To ensure that there is no ``placebo effect'' we also attempted to train the network after randomizing the input-output mappings. As expected, the training did not converge and the $R^2$ scores were close to zero or negative.

In Figure \ref{fig:reg} we plot a) the original census values and b) the predicted output of the direct regression model for some asset indicators on choropleth maps for the six states.  In Figure \ref{fig:regwb} we zoom into West Bengal for the ``water treated'' indicator. The salt and pepper noise in  the census data  indicates random errors in the data collection process. As is evident, the noise is smoothed out in the regression output which has more geo-spatial consistency. Similar effects are observed for all the asset indicators. 

Smoothing the raw census data over local neighbourhood of villages can also reduce the salt and pepper noise, but averaging out the error using the non-linear regression model learnt from over 2,00,000 villages spread across six states is much richer.

A significant deviation of the original census data from the predicted value for a village would indicate error in the census data. Thus,  not only do we obtain more accurate prediction, but we also get a census validation tool as a by-product.



\begin{figure}[h!]
\begin{center}
\begin{tabular}{c}
    \includegraphics[width=0.75\textwidth]{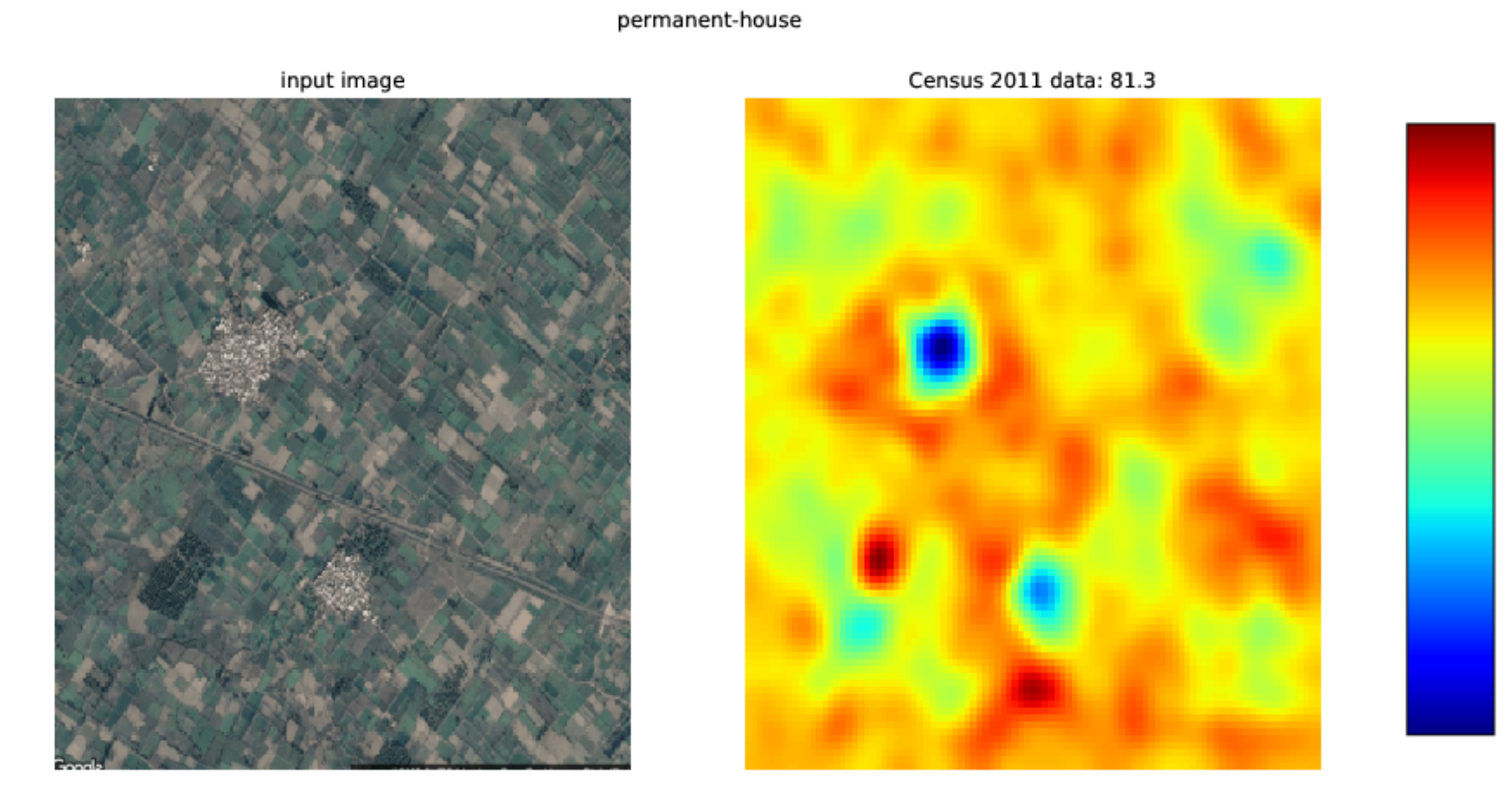} \\
     \includegraphics[width=0.75\textwidth]{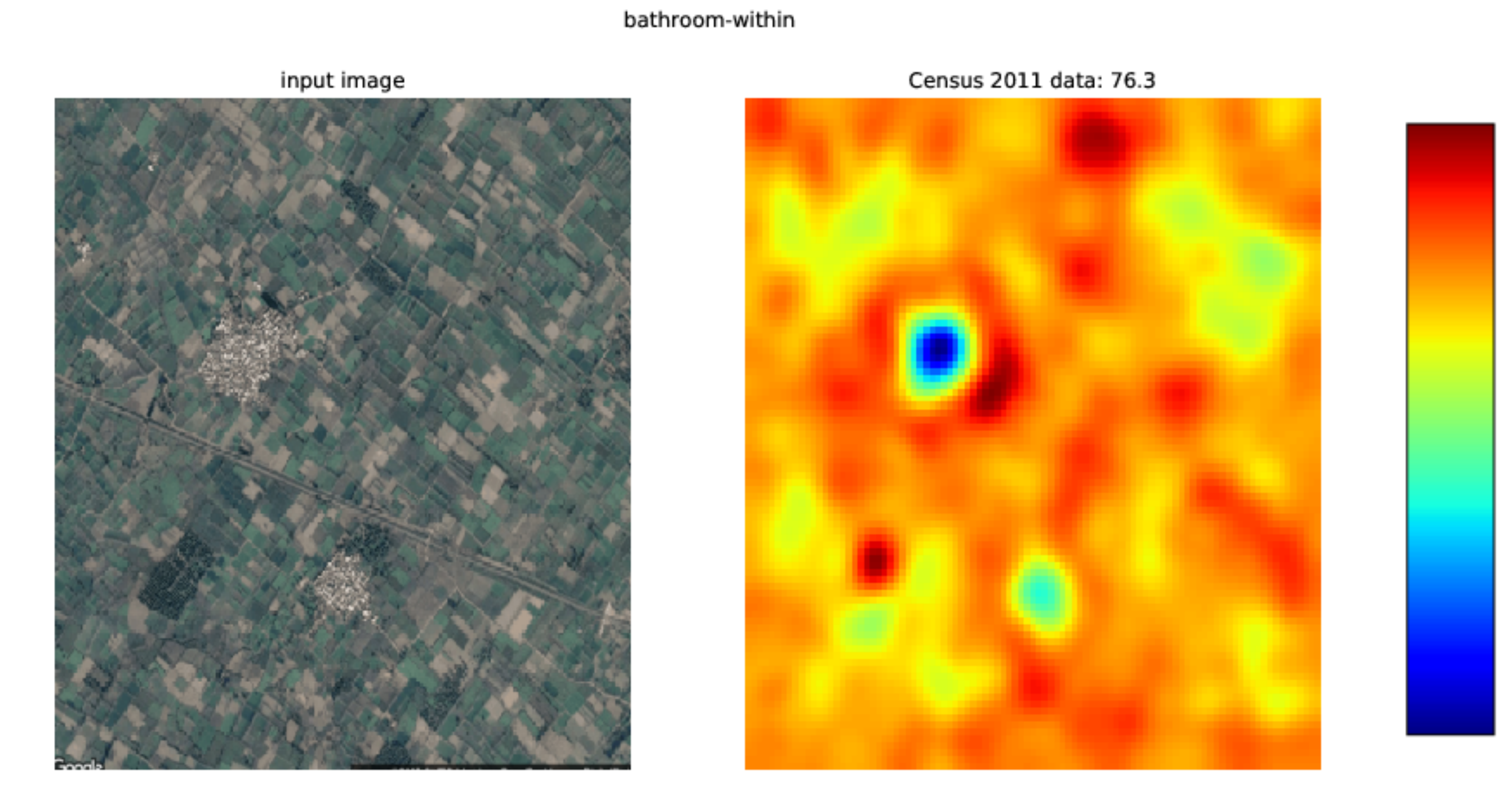} \\
      \includegraphics[width=0.75\textwidth]{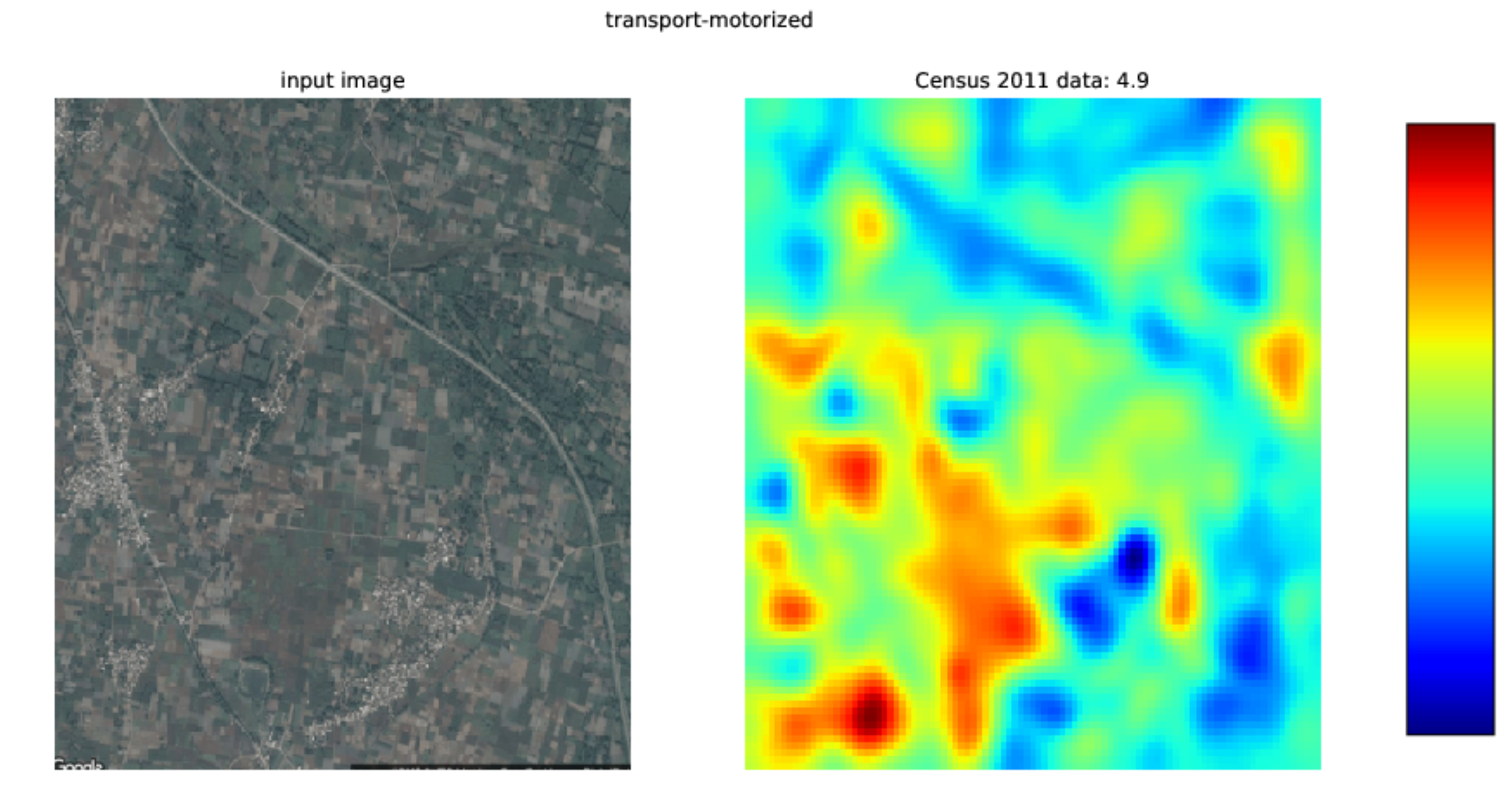}
\end{tabular}
    \caption{We plot heatmaps to understand what parts of an image may be responsible for regression of an asset indicator. The built-up and inhabited areas clearly show a dip when occluded for the ``permanent house'' and ``bathroom within'' indicators. Motorized transport are likely to be observed in developed villages with good roads as clearly illustrated by a heatmap for the ``transport motorized'' indicator.}
    \label{fig:heatmaps}
\end{center}
\end{figure}

\subsection{Salient image features for prediction}

As we have mentioned in Section \ref{sec:ML}, the interpretability of a deep CNN model is low. The features measured from images that are responsible for accurate regression are distributed in the weights of the CNN making it hard to decipher what exactly contributed to the accuracy. This, in turn, makes any causal analysis difficult. However, there have been some promising recent research \citep{DBLP:journals/corr/SimonyanVZ13,Mahendran:2016:VDC:2995936.2995953,DBLP:journals/corr/abs-1708-01785,DBLP:journals/corr/WeiZTF15,DBLP:journals/corr/ZeilerF13} towards interpreting deep CNNs which may eventually lead to better interpretation of CNN based models.

We use the method suggested in \citep{DBLP:journals/corr/ZeilerF13} in some sample images to understand what makes the regression possible. We slide a $16 \times 16$  occluder object over the images to investigate which parts of an image is responsible for the regression accuracy and plot this as a two dimensional heatmap. We show some sample results in Figure \ref{fig:heatmaps}. A sharp drop in the heatmap value (indicated in blue) when a region is occluded indicates that the region is significant for the regression outcome. Note that the regression model automatically learns what to control for in the images.

\section{Transfer learning of other socio-economic parameters}
\label{sec:transfer}

\begin{table}[h!]
\centering
\caption{Cross validated (out-of-sample) $R^2$ scores for the three transfer learning models.}
\label{table:transfer}
\begin{tabular}{ |p{4cm}|p{3cm}|p{3cm}|p{3cm}| }
\hline
 & from original asset data & from the night light model & from the direct regression model\\
\hline
Indicator & & & \\
\hline
 Literacy rate & 0.338 & 0.170 & \textbf{0.340} \\
 \hline
Scheduled caste (SC) percentage & 0.033 & 0.032 & \textbf{0.102} \\
 \hline
Scheduled Tribe (ST) percentage & 0.303 & 0.323 & \textbf{0.531} \\
 \hline
Percentage of working population & 0.181 & 0.125 & \textbf{0.194} \\
\hline
Overall variance weighted R2 score & 0.19 & 0.176 & \textbf{0.314} \\
\hline
\end{tabular}
\end{table}

\begin{figure}[h!]
\begin{tabular}{cc}
\includegraphics[width=3.0in]{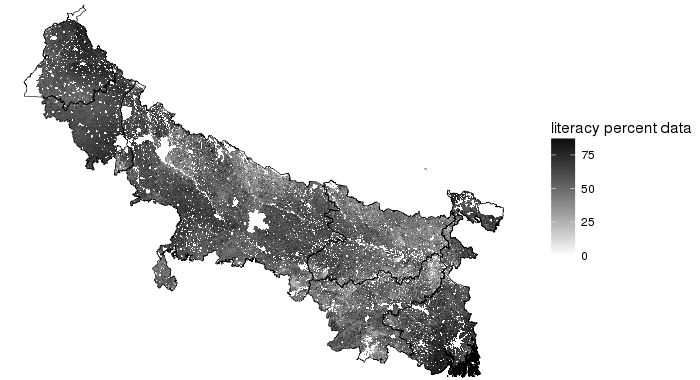} & \includegraphics[width=3.0in]{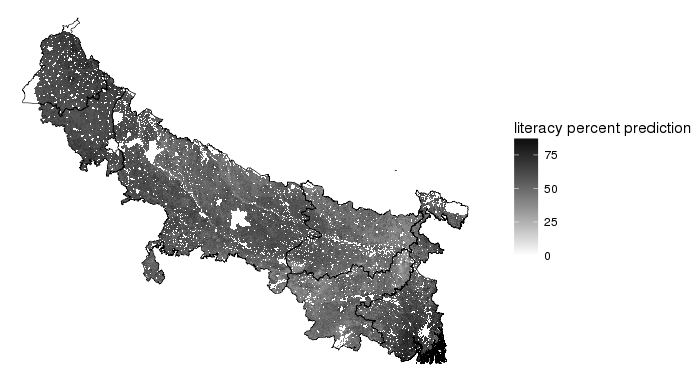}  \\
\includegraphics[width=3.0in]{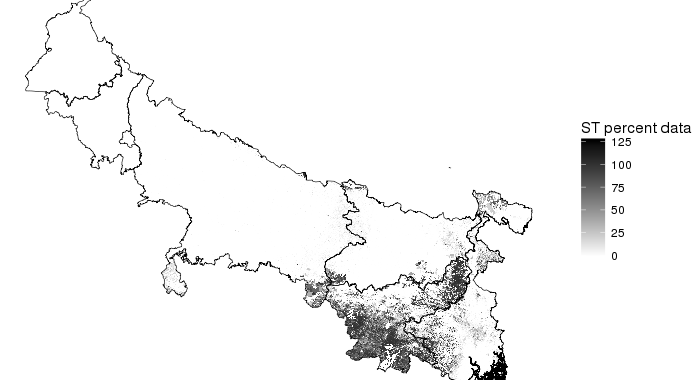} & \includegraphics[width=3.0in]{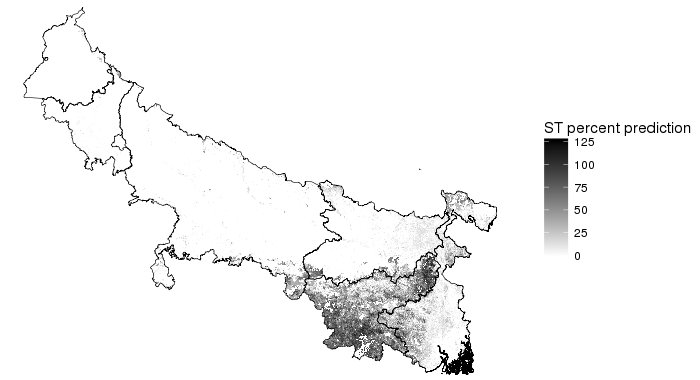}  \\
\includegraphics[width=3.0in]{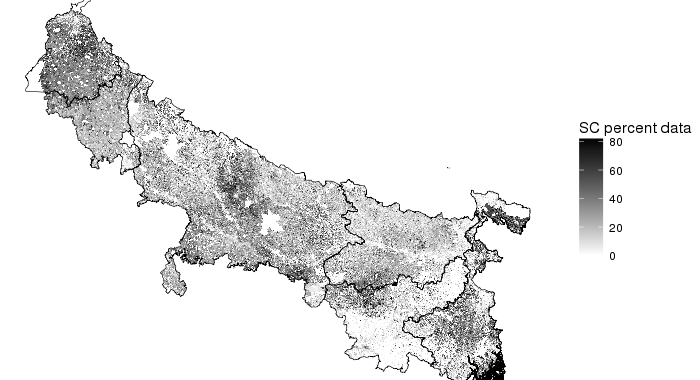} & \includegraphics[width=3.0in]{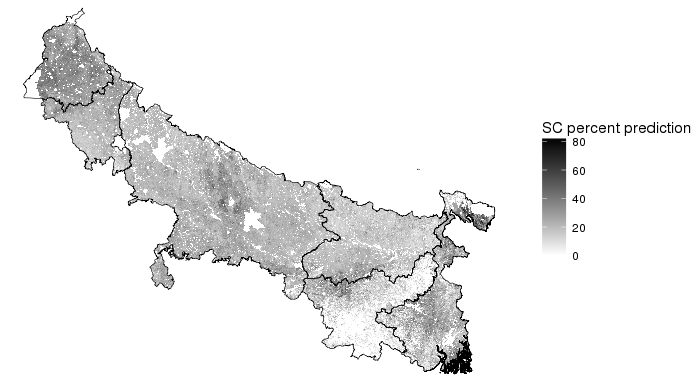}  \\
\includegraphics[width=3.0in]{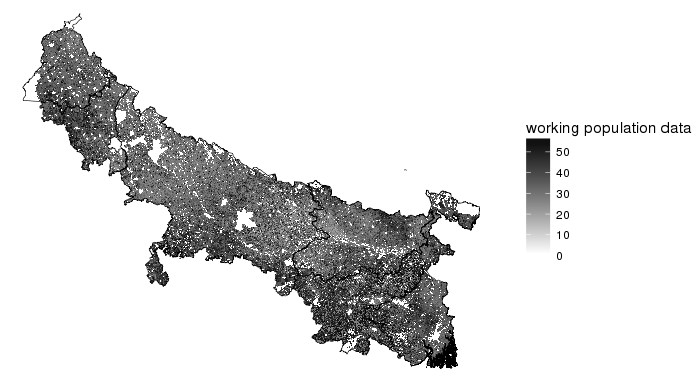} & \includegraphics[width=3.0in]{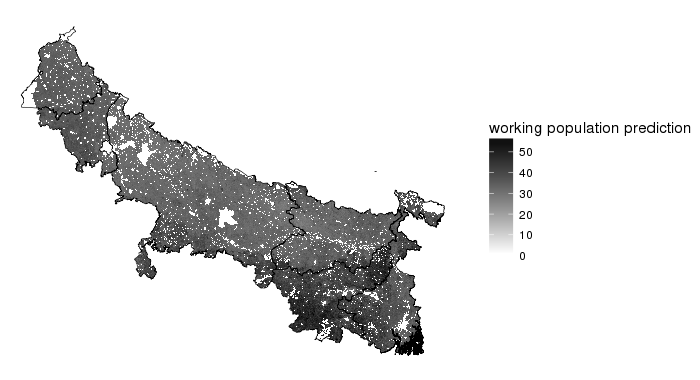}  \\
(a) Census data & (b) Prediction
\end{tabular}
\caption{Census data and transfer learning prediction from the direct regression model for percentage of `literate', `ST, `SC' and  `working' population in each village.}
\label{fig:transferkishen}
\end{figure}

Intuitively, the asset indicators may be correlated, either directly, or in some kernel space through nonlinear mappings, with the observable features in the daytime satellite images like proportion of built-up area, road area and road types, density and type of housing, water bodies, forest cover and green areas etc. This is confirmed by the regression models presented in Section \ref{sec:regression}. In this section we investigate whether the direct regression model trained to predict the asset indicators can be used for transfer learning \citep{5288526} of other socio-economic indicators for literacy, health and demographics which are not directly related to what can be measured from satellite imagery. 

In Table \ref{table:transfer} we present the leave-one-out cross validation $R^2$ scores of three fully connected  two layered neural network models with rectified linear activations \citep{Goodfellow-et-al-2016} trained to predict a few socio-economic indicators obtained from Census 2011 \citep{census} (Population Enumeration) data in all villages in the six states using the following as input:
\begin{enumerate}[label=(\subscript{M}{{\arabic*}})]
\item the asset vector computed from the raw census data.
\item the 4096 dimensional feature vector obtained from the last layer of the night light model of Section \ref{sec:regnightlight}.
\item \label{enum:direct} the 4096 dimensional feature vector obtained from  the last layer of the direct regression model of  Section \ref{sec:regdirect}.
\end{enumerate}

\begin{table}[h!]
\centering
\caption{Socio-economic and health indicators from NFHS-4.}
\label{table:nfhs4desc}
\begin{tabular}{|p{1.5cm}|p{5cm}|p{6.5cm}|}
\hline
Indicator & NFHS-4 section & Description \\
\hline \hline
8-rural & Population and Household Profile (rural) & Households using improved sanitation facility (\%) \\ \hline
33-rural & Maternity Care (rural) & Mothers who had full antenatal care (\%) \\ \hline
35-rural & Maternity Care (rural) & Mothers who received postnatal care from a doctor/nurse/LHV/ANM/midwife/other health personnel within 2 days of delivery (\%) \\ \hline
41-rural & Delivery Care (rural) &  Institutional births in public facility (\%) \\ \hline
47-rural & Child Immunizations and Vitamin A Supplementation (rural) & Children age 12-23 months fully immunized (BCG, measles, and 3 doses each of  polio and DPT) (\%) \\ \hline
68-rural & Child Feeding Practices and Nutritional Status of Children (rural) &  Children under 5 years who are stunted (height-for-age) (\%) \\ \hline
69-rural & Child Feeding Practices and Nutritional Status of Children (rural) & Children under 5 years who are wasted (weight-for-height) (\%) \\ \hline
70-rural & Child Feeding Practices and Nutritional Status of Children (rural) &  Children under 5 years who are severely wasted (weight-for-height) (\%) \\ \hline
\end{tabular}
\end{table}

Clearly, transfer learning from the direct regression model \ref{enum:direct} outperforms the other two. This also demonstrates  that the predicted asset indicators are superior information sources than what is computed from the raw census data. In Figure \ref{fig:transferkishen} we present the  choropleth maps for transfer learning from the direct regression model.

We also try transfer learning of some socio-economic and health indicators from the predicted asset vectors using the Census 2011 \citep{census} and NFHS-4 survey data \citep{nfhs4}. Some of the NFHS-4 indicators are described in Table \ref{table:nfhs4desc}. The ``education levels'' data in Census 2011 (Population Enumeration Data - C-08 Educational Level By Age And Sex For Population Age 7 And Above (Total, SC/ST) (India \& States/UTs-District Level)), and the NFHS-4 data are available at a lower granularity of district level, and there are a total of 192 districts in the six north Indian states. We consider only `rural' part of the districts. We ignore 10 districts which are completely urban.

For transfer learning of these indicators we aggregate the asset model  \ref{enum:direct} regression output  for all villages in a district by averaging, and train a neural network with a single fully connected layer with rectified linear activation \citep{Goodfellow-et-al-2016} to do a regression of the NFHS-4 and ``education level'' indicators from the 16 dimensional input. We do a 5 part split for the 182 `rural' districts and train using leave-one-out cross validation. The $R^2$ scores that we report is on the validation set for the best performing model.

\begin{figure}[h!]
\begin{tabular}{cc}
\includegraphics[width=3.0in]{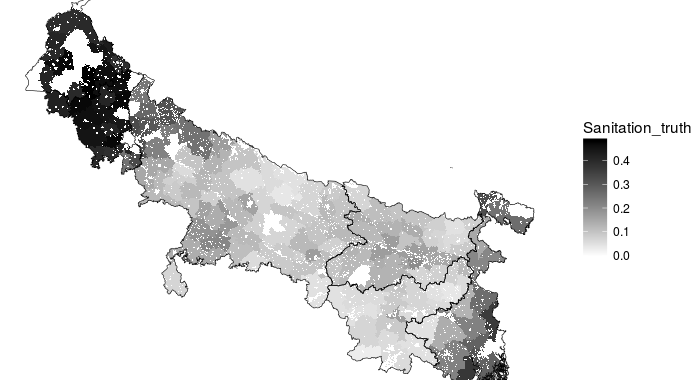} & \includegraphics[width=3.0in]{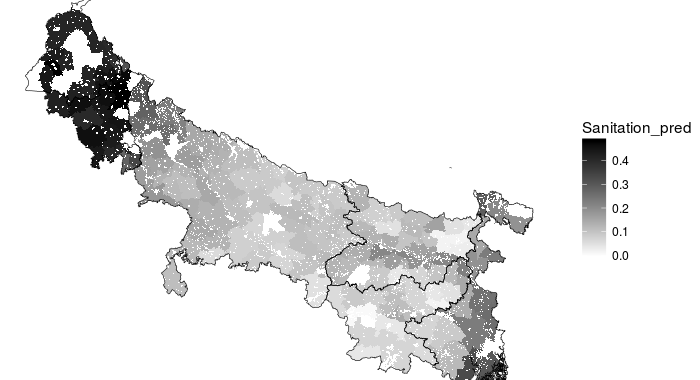}  \\
\includegraphics[width=3.0in]{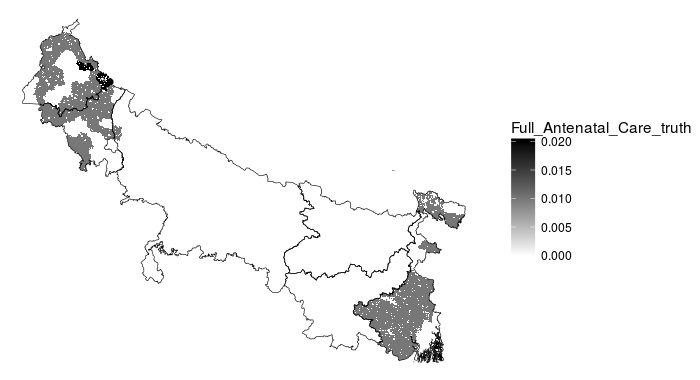} & \includegraphics[width=3.0in]{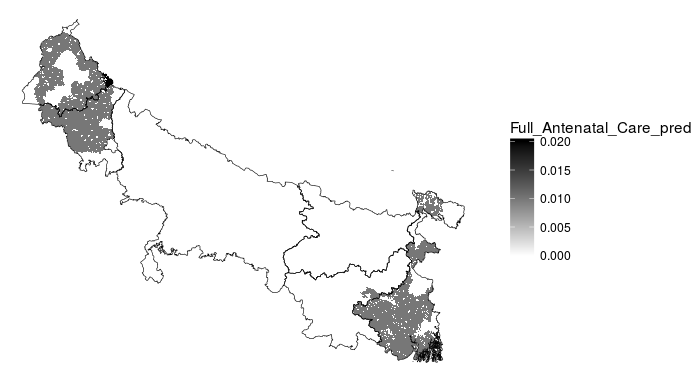}  \\
\includegraphics[width=3.0in]{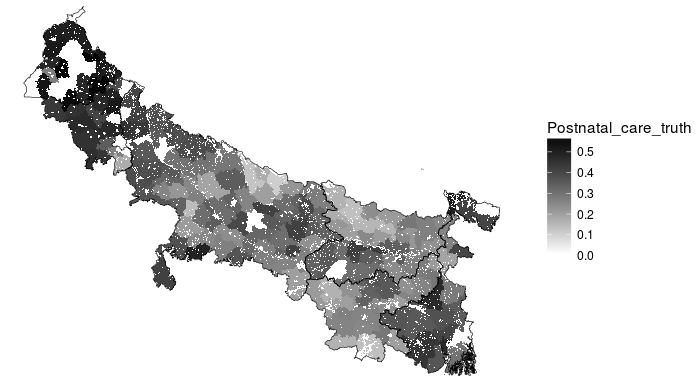} & \includegraphics[width=3.0in]{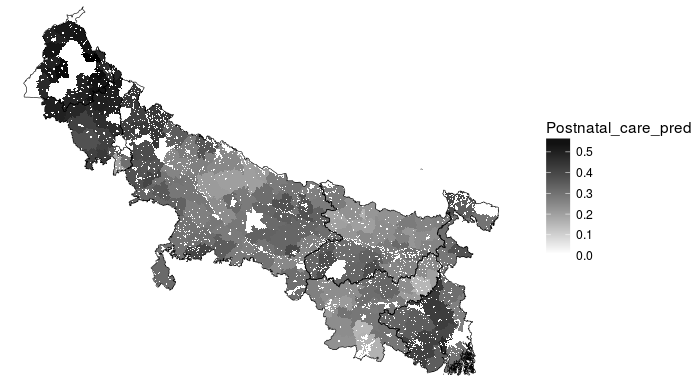}  \\
\includegraphics[width=3.0in]{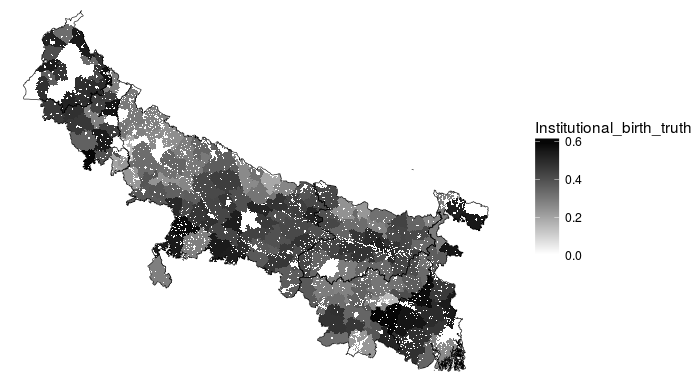} & \includegraphics[width=3.0in]{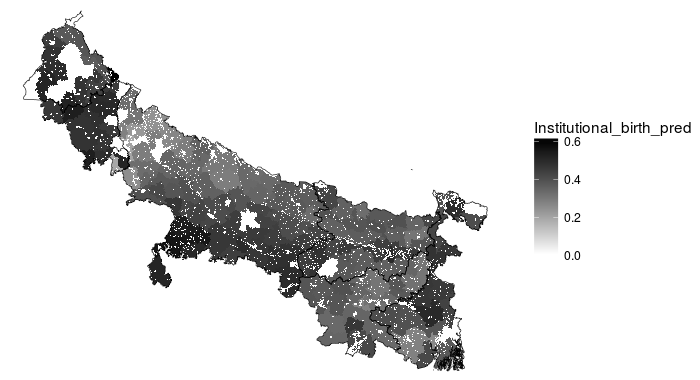}  \\
(a) NFHS-4 data & (b) Prediction
\end{tabular}
\caption{NFHS-4 data and transfer learning prediction for ``8-rural'', ``33-rural'', ``35-rural'' and ``41-rural'' indicators. The out-of-sample $R^2$ scores for regression were 0.73, 0.57, 0.45 and -0.6 respectively. The grayscale indicates proportions.}
\label{fig:transfera}
\end{figure}

\begin{figure}[h!]
\begin{tabular}{cc}
\includegraphics[width=3.0in]{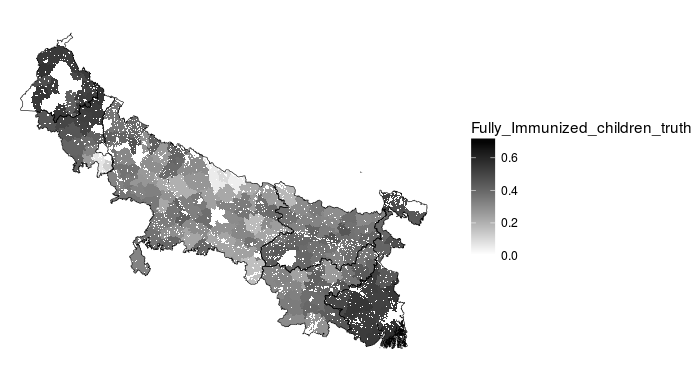} & \includegraphics[width=3.0in]{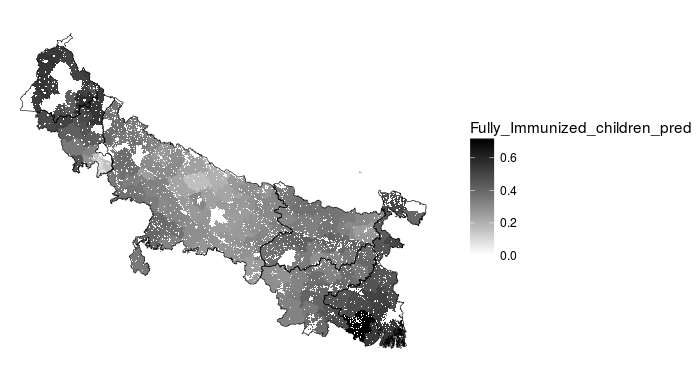}  \\
\includegraphics[width=3.0in]{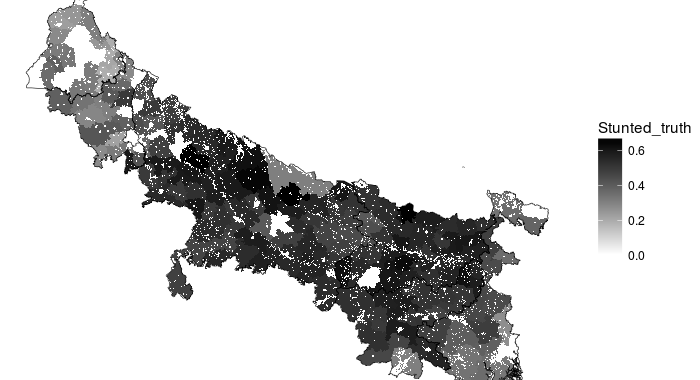} & \includegraphics[width=3.0in]{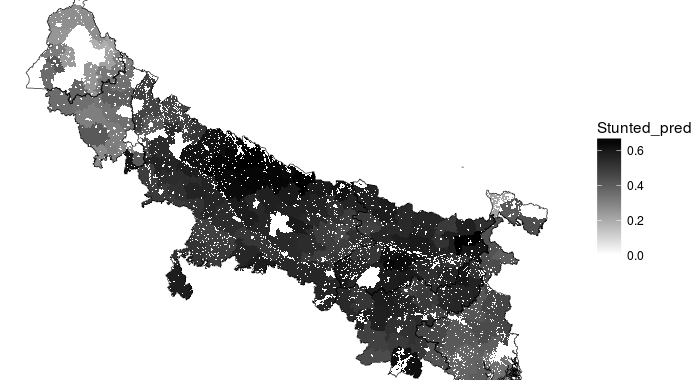}  \\
\includegraphics[width=3.0in]{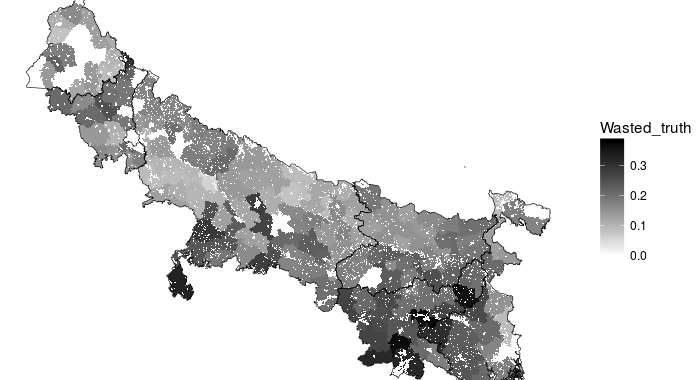} & \includegraphics[width=3.0in]{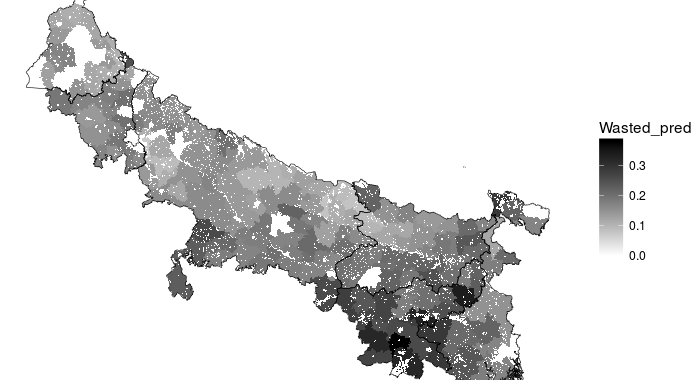}  \\
\includegraphics[width=3.0in]{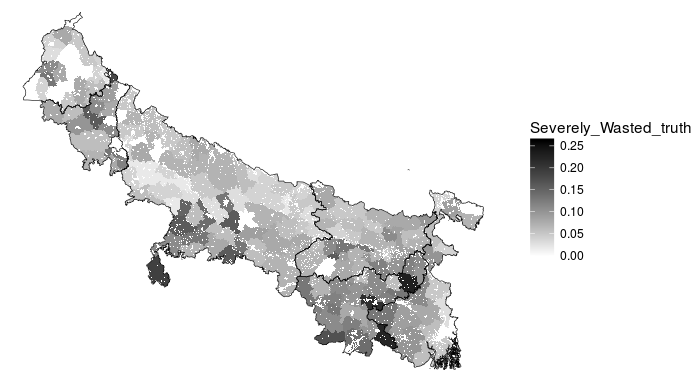} & \includegraphics[width=3.0in]{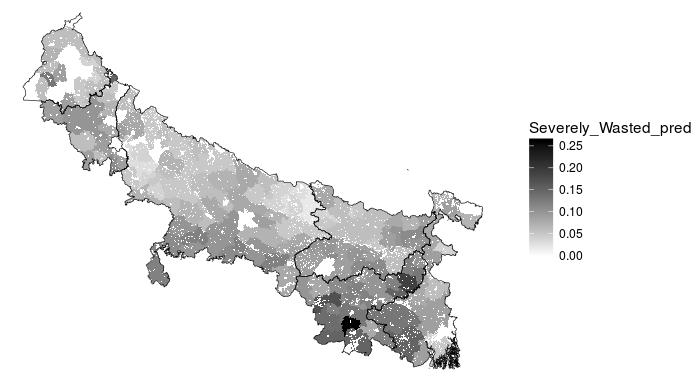}  \\
(a) NFHS-4 data & (b) Prediction
\end{tabular}
\caption{NFHS-4 data and transfer learning prediction for ``47-rural'', ``68-rural'', ``69-rural'' and ``70-rural'' indicators. The out-of-sample $R^2$ scores for regression were 0.61, 0.61, -0.45, -0.45 respectively. The grayscale indicates proportions.}
\label{fig:transferb}
\end{figure}

\begin{figure}[h!]
\begin{tabular}{cc}
\includegraphics[width=3.0in]{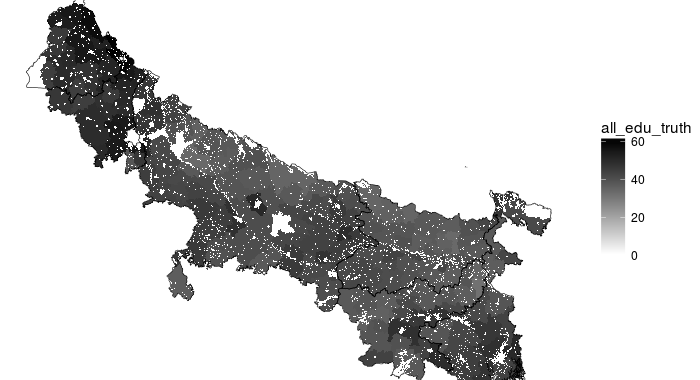} & \includegraphics[width=3.0in]{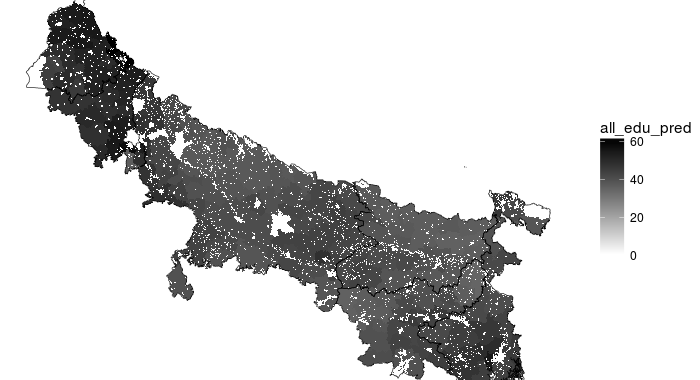}  \\
(a) Census data & (b) Prediction
\end{tabular}
\caption{Census data and transfer learning prediction for ``average education level''. The out-of-sample $R^2$ score for regression was 0.54. The grayscale indicates percentages.}
\label{fig:transferedu}
\end{figure}

\begin{figure}[h!]
\begin{center}
\includegraphics[width=0.75\textwidth]{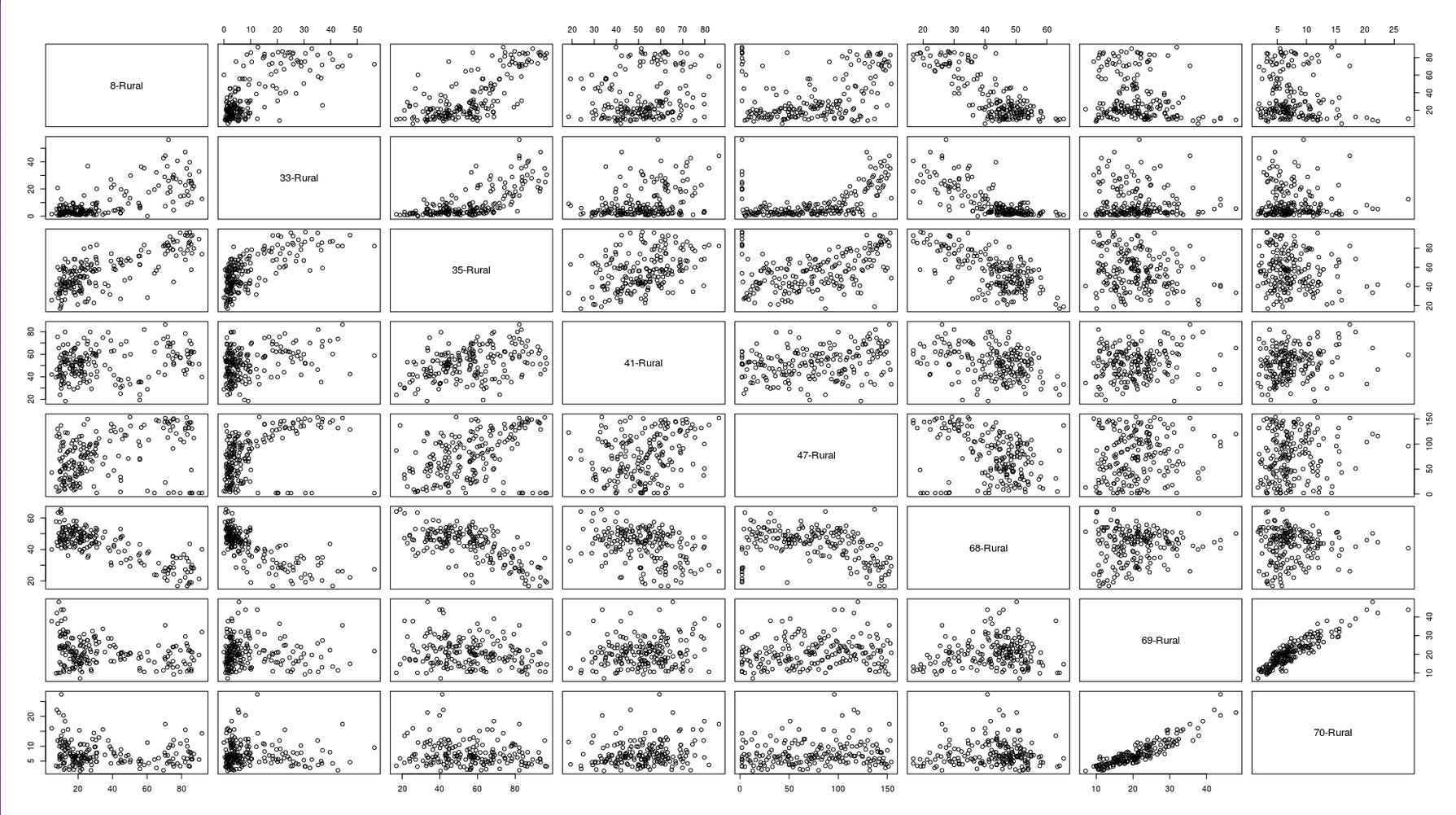}
\end{center}
\caption{Pair-wise  scatter plots of the NFHS-4 indicators computed over 182 districts.}
\label{fig:nfhs4corr}
\end{figure}

\begin{figure}[h!]
\begin{center}
\includegraphics[width=0.5\textwidth]{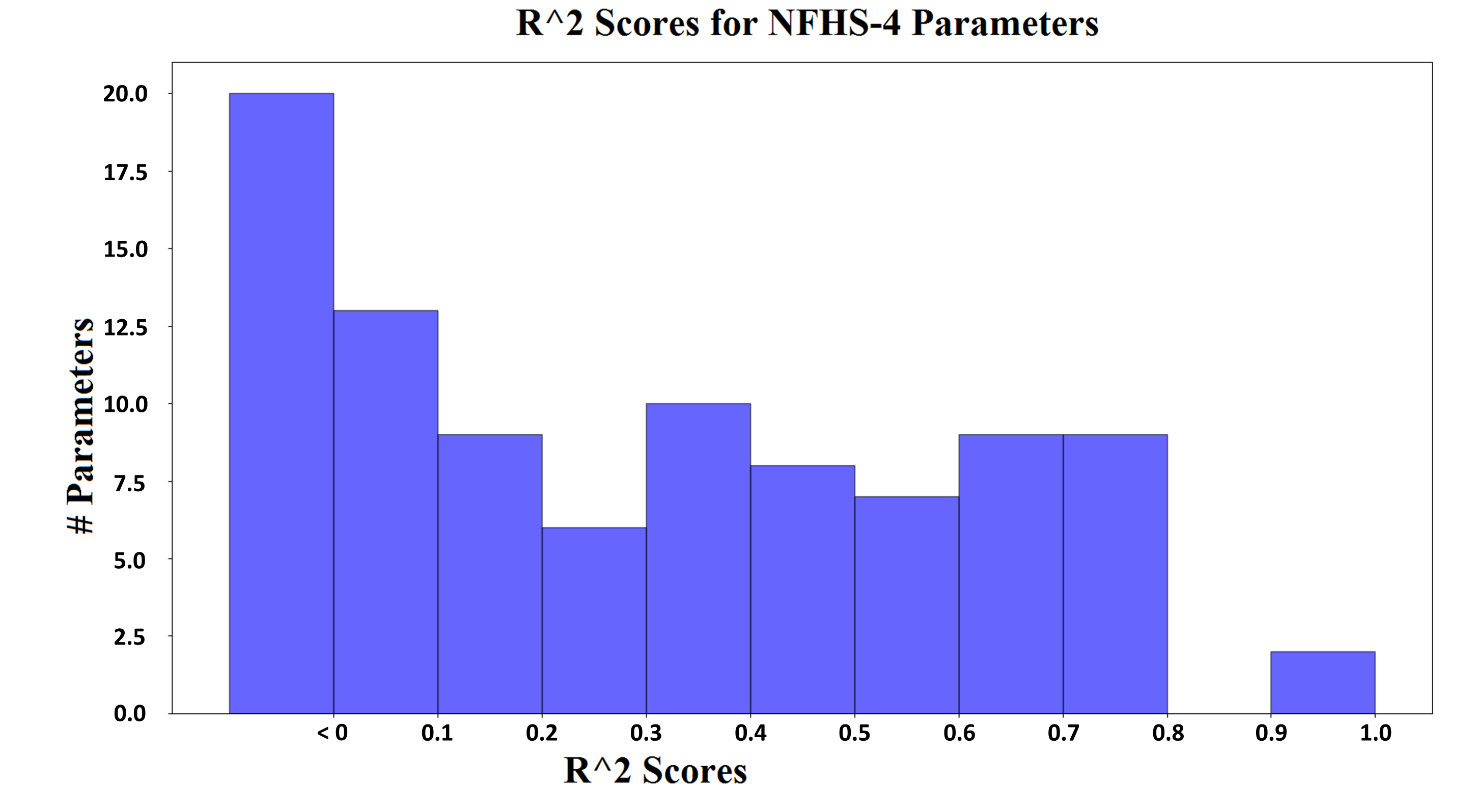}
\end{center}
\caption{Histogram of out-of-sample $R^2$ scores of regression of all 93 `rural' parameters of NFHS-4. See \href{http://web.iitd.ac.in/~suban/satellite/nfhs4_r2_scores.html}{http://web.iitd.ac.in/$\sim$suban/satellite/nfhs4\_r2\_scores.html} for the $R^2$ scores and \href{http://web.iitd.ac.in/~suban/satellite/nfhs4/}{http://web.iitd.ac.in/$\sim$suban/satellite/nfhs4/} for the choropleths.} 
\label{fig:nfhs4hist}

\end{figure}

 In  Figures \ref{fig:transfera}, \ref{fig:transferb} and \ref{fig:transferedu}  we show the state-wise choropleth maps for the ground truth and  prediction of the NFHS-4 and ``education level'' indicators. We also indicate the regression $R^2$ scores for a cross validation set.
 
 In Figure \ref{fig:nfhs4corr} we show the pair-wise scatter plots of the NFHS-4 indicators computed over all 182 districts.  As can be noted that all the NFHS-4 indicators apart from ``institutional birth'' and ``wasting'' can be accurately predicted from regression output of the asset model, even though some of them are poorly correlated among each other. The highly nonlinear functional relationships are surprisingly well captured by the transfer learning. 
 
We also try transfer learning of all the 93 `rural' parameters in NFHS-4. In Figure \ref{fig:nfhs4hist}, we show the histogram of $R^2$ scores. Even though the NFHS-4 parameters are  not intuitively related to what can be observed from satellite images, we obtain reasonable prediction accuracy for over 75\% of them. As we demonstrate in Section \ref{sec:stunting}, such predictions can be used to remove possible omitted variable biases and predict endogenous variables at the first stage of a instrumental regression experiment.

\section{Monitoring development over time}
\label{sec:temporal}

\begin{figure}[h!]
\centering
\begin{tabular}{ccc}
\includegraphics[width=2in]{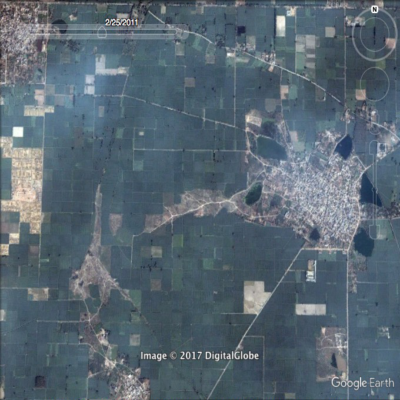} & \includegraphics[width=2in]{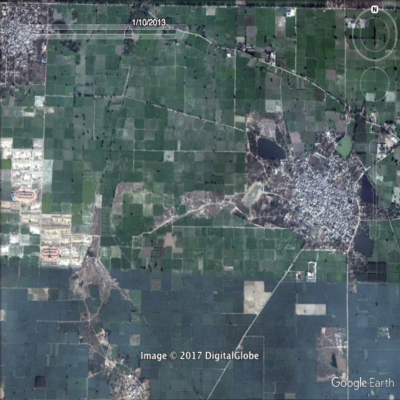} & \includegraphics[width=2in]{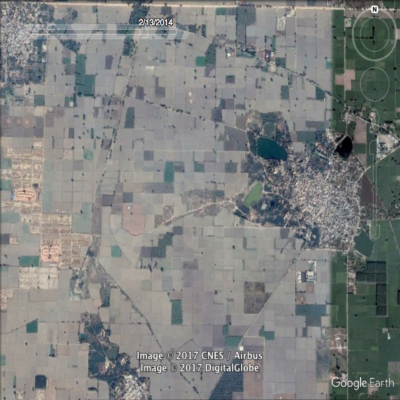} \\
 Feb 25, 2011 & Jan 10, 2013 & Feb 13, 2014 \\
\includegraphics[width=2in]{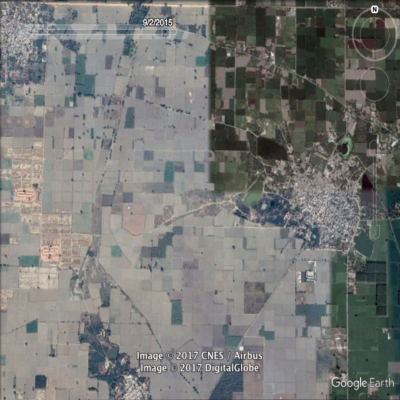} & \includegraphics[width=2in]{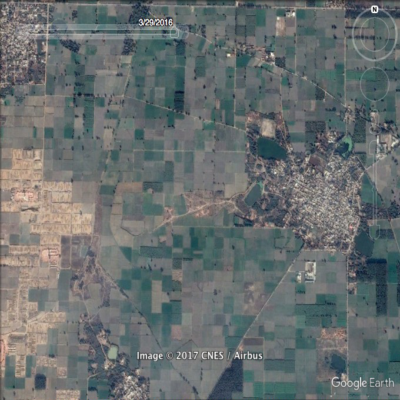} & \includegraphics[width=2in]{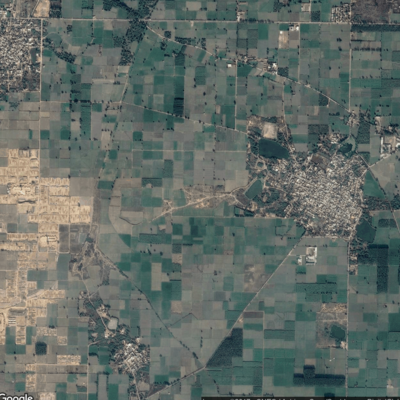} \\
Feb 9, 2015 & Mar 29, 2016 & Feb 2017 (mosaic) 
\end{tabular}
\caption{Images of Bhadana village, Sonipat tehsil, Haryana state captured at different times between 2011-2017.}
\label{fig:temporal}
\end{figure}

\begin{figure}[h!]
	\centering
    \includegraphics[width=1.0\textwidth,height = 0.8\textheight]{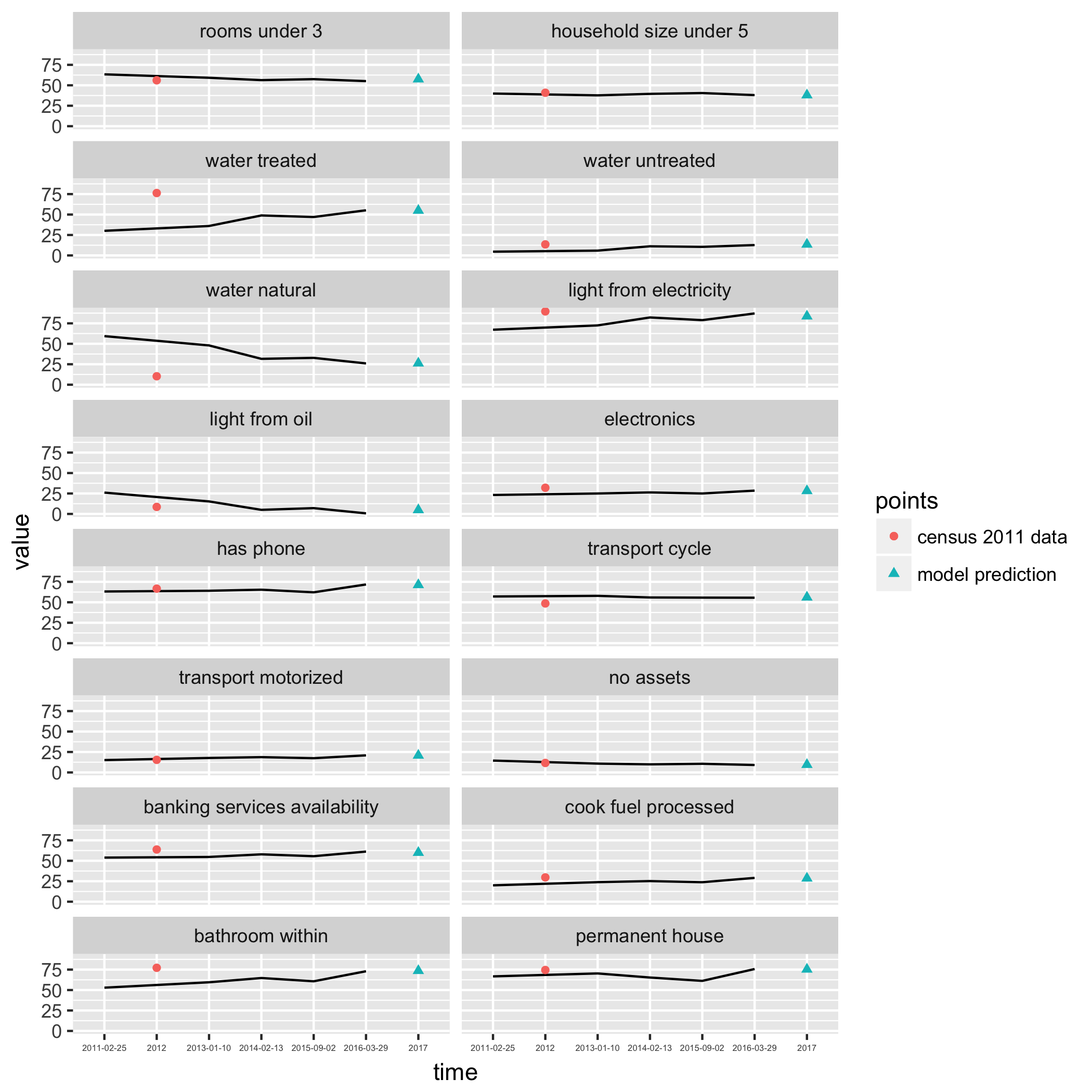}
    \caption{The evolution of asset indicators over time}
    \label{fig:evolution}
\end{figure}

Despite the fact that we use disparate cross-sectional data - daytime satellite images captured in 2017 and census data of 2011 - to train the regression model, we obtain reasonably high regression accuracy. This can be attributed to the fact that the large number of villages (over 2,00,000) that we use to train the model are at varying stages of economic development, and their collective  diversity is rich enough to represent the characteristics  of economic development spanning several years. This provides us with an opportunity to use the static regression model trained with cross-sectional data to monitor the temporal evolution of a village.


In Figure \ref{fig:temporal} we show images (downloaded manually) of a village captured at different times between 2011 and 2017. In Figure \ref{fig:evolution} we present the regression output of our asset indicators for these images. Despite the fact that the cross-sectional regression of Census 2011 indicators from the 2017 image is not very accurate for this village, indicating that the Census 2011 data for this village is not accurate, the smooth and near monotonic development of the village through the years is evident from the predicted values. We have observed similar smooth predictions for over 35 villages. Ground level intermediate surveys are required for a comprehensive validation of the temporal predictions.

\section{Spatial discontinuities in regression output}
\label{sec:rd}

\begin{figure}[ht]
\begin{tabular}{cc}
\includegraphics[width=3.0in]{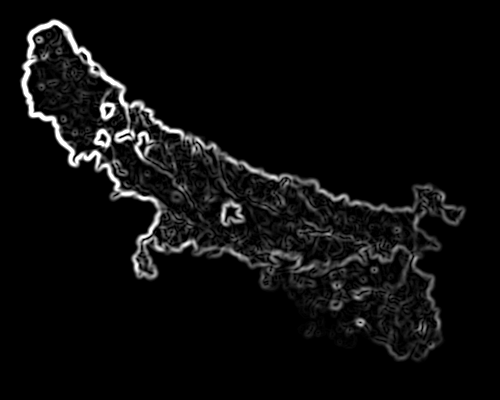} & \includegraphics[width=3.0in]{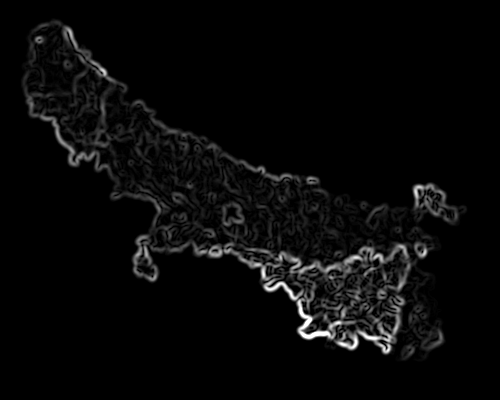}  \\
(a) bathroom.within & (b) water.untreated  \\
\end{tabular}
\caption{Sharp spatial gradients in predicted values of  bathroom.within and  water.untreated}
\label{fig:rd}
\end{figure}

We argue that spatial discontinuities or sharp spatial gradients in the regression output of a development indicator require special attention. Such high gradients can be computed by carrying out edge detection \citep{gonzalez2009digital} in the regression output choropleth maps - see Figure \ref{fig:rd} for an example. Edge detection involves computing the image gradients using finite differences and thresholding the high gradient image locations. An edge in the regression output indicates that geographically neighbouring villages have dissimilar economic development outcomes. Often such edges can occur at village-forest or city-village boundaries and can be easily explained away. Otherwise, if two neighbouring villages are similar in all other aspects, then  an edge in the regression output  may suggest that the  villages have been subjected to different policy interventions or there may be other socio-economic anomalies. All such edge points require special attention and investigation to find  causal explanations. Indeed, not surprisingly, the edges in  Figure \ref{fig:rd} often coincide with state or district boundaries where neighbouring villages belonging to different states  or districts have been subjected to different policy interventions. 

\section{Using predicted variables for regression: a case study to understand stunting}
\label{sec:stunting}

Finally, we do a case study of regression analysis using predicted variables to understand the determinants of the disturbingly high rate of {\em stunting} (see second row of Figure ~\ref{fig:transferb}) in the north Indian states. The determinants of stunting are not yet completely understood \citep{stuntingepw}, and we analyse the robustness of a recent regression study carried out by \cite{Spears2013}.

Identification with observational data  in econometrics crucially depends on to what extent one can maintain the assumption of exogeneity of explanatory variables. One of the main reasons for violation of exogeneity arises from inability to control for omitted sources of variation which may be correlated with the included explanatory variables. Finding a good proxy for an omitted variable is a difficult task, either because of missing variables in the data, or because of missing observations,  or both. Missing observations  usually require extrapolation or interpolation  based  projections, whereas missing variables require finding a suitable proxy from other data sources. We show that  machine learning based predictions can not only account for missing observations but can also be useful for generation of  proxy variables and correcting for omitted variable biases. Our predictive tool can also correct for attenuation biases due to measurement errors such as in Census and account for the classical ``error in variables'' problem.

\cite{Spears2013} hypothesise that prevalence of open defecation is an important correlate of stunting rates in India. The unit of their analysis is a non-representative sample of 112 districts out of over 600 districts in India. They use four different data sets, spread over 2005 to 2011,  to collect the relevant variables. The outcome variable (stunting rates) is from a survey in 2010-11, the open defecation rates are collected from the 2011 Census report, infant mortality rates are  from Annual Health Survey 2010-2011  and  the consumption and calorie data are collected from 2005 National Sample Survey. The district level stunting rate is regressed on $\log$ of open defecation rate, controlling for several other covariates like urban population (quadratic), $\log$ of monthly per capita consumption expenditure (MPCE), calories per capita, cereal calories per capita, household size, overall literacy rate and female literacy rate. Their main conclusion  is that districts with more open defecation have more stunted children and that it is robust with respect to multiple controls. With the full set of controls, a 10\% increase in open defecation rate is associated with 0.7\% increase in the stunting rate. This association fades away somewhat when they also control for infant mortality rate (IMR), which is endogeneous in the specification (see Table ~\ref{table:distSGC}). 

\begin{table}[h!]
{\tiny
\centering
\caption{Comparison of our models with \citep{Spears2013}.  Note that all but the calories are proportions. The calorie values range from 1000 to 2500.}
\label{table:distSGC}
\begin{tabular}{p{3cm}rrrr|rrr}
\toprule
 & \multicolumn{4}{c|}{Predicted data} & \multicolumn{3}{c}{\citep{Spears2013} (SGC)} \\
\midrule
 & m1 & m2 & m3 & m4 & m2-SGC & m3-SGC & m4-SGC \\
 & b/se & b/se & b/se & b/se & b/se & b/se & b/se \\
 \midrule
ln(opendefecation) & 7.404 & 3.621 & 1.326 & 1.374 & 7.969 & 8.628 & 7.082 \\
 & (0.66) & (0.68) & (0.60) & (0.67) & (1.597) & (2.472) & (2.803) \\
ln(mpce) &  & -9.266 & -0.740 & -1.623 &  & 8.765 & 8.103 \\
 &  & (1.55) & (1.44) & (1.34) &  & (7.816) & (6.986) \\
Calories per capita &  & -0.005 & -0.000 & 0.000 &  & -0.0119 & -0.00466 \\
 &  & (0.00) & (0.00) & (0.00) &  & (0.0102) & (0.00907) \\
householdsizeunder5 &  & -0.006 & -0.271 & -0.230 \\
 &  & (0.07) & (0.06) & (0.08) \\
literacy-rate &  &  & -0.716 & -0.448 &  &  & -0.810 \\
 &  &  & (0.12) & (0.14) &  &  & (0.450) \\
women-lit &  &  & -0.151 & -0.022 &  &  & 0.335 \\
 &  &  & (0.05) & (0.05) &  &  & (0.461) \\
mom-folic &  &  &  & 0.047 \\
 &  &  &  & (0.05) \\
women-sec-edu &  &  &  & -0.174 \\
 &  &  &  & (0.08) \\
mom-full-ant-care &  &  &  & -0.021 \\
 &  &  &  & (0.07) \\
ceasarean-birth &  &  &  & -0.019 \\
 &  &  &  & (0.08) \\
children-vitA &  &  &  & -0.095 \\
 &  &  &  & (0.03) \\
women-bmi-below-norm &  &  &  & 0.192 \\
 &  &  &  & (0.10) \\
clean-fuel &  &  &  & -0.026 \\
 &  &  &  & (0.05) \\
percent urban  &  &  &  &  & 0.216  & 0.128 & 0.246 \\
 &  &  &  &  & (0.218) & (0.216) & (0.173) \\
percent urban squared &  &  &  &  & -0.00504 & -0.00480 & -0.00512 \\
 &  &  &  &  & (0.00290) & (0.00274) & (0.00229) \\
Cereal Calories per capita &  & 0.020 & 0.011 & 0.006 &  & 0.00235 & -0.00476 \\
 &  & (0.00) & (0.00) & (0.00) &  & (0.00816) & (0.00782) \\
household size &  &  &  &  &  & 1.470 & 1.595 \\
 &  &  &  &  &  & (1.077) & (0.852) \\
Constant & 12.827 & 80.076 & 77.795 & 74.627 & 22.09 & 22.28 & 15.80 \\
 & (2.62) & (14.17) & (11.09) & (10.77) & (7.647) & (46.13) & (43.93) \\
 \midrule
R-squared & 0.411 & 0.733 & 0.839 & 0.869 & 0.453 & 0.484 & 0.586 \\
N & 178 & 178 & 178 & 178 & 112 & 112 & 112 \\
\bottomrule
\end{tabular}
}
\end{table}

We replicate this regression using our predicted values for all the districts in the six north Indian states. We leave out a few districts where the predicted rate for the outcome variable or that of an important control variable turns out to be negative (this can happen in a predictive model).  The sample districts in our study are different from that of \citep{Spears2013}. Our constructed sample of district level variables are only for the rural areas, therefore our first model is comparable with the model 2 of \citep{Spears2013} (See Table ~\ref{table:distSGC}). We obtain our district level predictions from transfer learning of the NFHS-4 and ``education level'' indicators described in Section \ref{sec:transfer}.

\begin{table}[h!]
{\tiny
\centering
 \caption{Comparison of models with predicted data and  actual data}
 \label{Table:predictedvsactual}
\begin{tabular}{p{2.5cm}rrrr|rrrr}
\toprule
 & \multicolumn{4}{c|}{Predicted data} & \multicolumn{4}{c}{Actual data} \\
\midrule
 & m1 & m2 & m3 & m4 & m1 & m2 & m3 & m4 \\
 & b/se & b/se & b/se & b/se & b/se & b/se & b/se & b/se \\
  \midrule
ln(opendefecation) & 7.404 & 3.621 & 1.326 & 1.374 & 8.437 & 5.325 & 2.291 & 1.873 \\
 & (0.66) & (0.68) & (0.60) & (0.67) & (0.68) & (0.76) & (0.71) & (0.90) \\
ln(mpce) &  & -9.266 & -0.740 & -1.623 &  & -13.719 & -8.729 & -6.307 \\
 &  & (1.55) & (1.44) & (1.34) &  & (3.69) & (3.14) & (3.13) \\
kcal2 &  & -0.005 & -0.000 & 0.000 &  & -0.004 & -0.001 & -0.001 \\
 &  & (0.00) & (0.00) & (0.00) &  & (0.00) & (0.00) & (0.00) \\
kcal-cereal &  & 0.020 & 0.011 & 0.006 &  & 0.004 & -0.001 & -0.002 \\
 &  & (0.00) & (0.00) & (0.00) &  & (0.01) & (0.00) & (0.00) \\
householdsizeunder5 &  & -0.006 & -0.271 & -0.230 &  & -0.034 & -0.202 & -0.083 \\
 &  & (0.07) & (0.06) & (0.08) &  & (0.07) & (0.06) & (0.08) \\
literacy-rate &  &  & -0.716 & -0.448 &  &  & 0.016 & -0.022 \\
 &  &  & (0.12) & (0.14) &  &  & (0.08) & (0.08) \\
women-lit &  &  & -0.151 & -0.022 &  &  & -0.383 & -0.327 \\
 &  &  & (0.05) & (0.05) &  &  & (0.05) & (0.09) \\
mom-folic &  &  &  & 0.047 &  &  &  & -0.041 \\
 &  &  &  & (0.05) &  &  &  & (0.08) \\
women-sec-edu &  &  &  & -0.174 &  &  &  & 0.083 \\
 &  &  &  & (0.08) &  &  &  & (0.11) \\
mom-full-ant-care &  &  &  & -0.021 &  &  &  & -0.060 \\
 &  &  &  & (0.07) &  &  &  & (0.10) \\
ceasarean-birth &  &  &  & -0.019 &  &  &  & 0.028 \\
 &  &  &  & (0.08) &  &  &  & (0.09) \\
children-vitA &  &  &  & -0.095 &  &  &  & -0.084 \\
 &  &  &  & (0.03) &  &  &  & (0.03) \\
women-bmi-below-norm &  &  &  & 0.192 &  &  &  & 0.023 \\
 &  &  &  & (0.10) &  &  &  & (0.10) \\
clean-fuel &  &  &  & -0.026 &  &  &  & -0.087 \\
 &  &  &  & (0.05) &  &  &  & (0.06) \\
Constant & 12.827 & 80.076 & 77.795 & 74.627 & 9.133 & 124.091 & 127.715 & 112.289 \\
 & (2.62) & (14.17) & (11.09) & (10.77) & (2.70) & (25.38) & (20.99) & (21.49) \\
 \midrule
R-squared & 0.411 & 0.733 & 0.839 & 0.869 & 0.449 & 0.657 & 0.766 & 0.790 \\
N & 178 & 178 & 178 & 178 & 191 & 188 & 188 & 188 \\
\bottomrule
\end{tabular}
  }
\end{table}

The estimates are remarkably similar.  However, when we control for $\log$ of MPCE, total calorie, calorie from cereal and percentage of household size under 5, our estimates are quite different from that of \citep{Spears2013}. We should keep in mind that our consumption data (predicted) comes from 2011-12 NSSO survey \citep{nsso}, whereas  the consumption data for \citep{Spears2013} are quite disparate (2005).  A significant part of the variation in stunting is explained by overall economic condition of the households as measured by district level average MPCE.  As expected, higher calorie consumption is negatively associated with stunting rate whereas calorie from cereals is positively associated with stunting  rate. This is not surprising because given two households with similar levels of total calorie intake, the one with more calorie from cereals (nutritionally not so rich) will have higher stunting possibilities. We may extend the same logic to district  level, where indeed the variance of the total per capita calorie intake is low. Unlike in our model, model 3 in \citep{Spears2013} does not indicate any of the control factors to be significant, whereas open defecation remains very significant and the estimated value does not drop. In the third specification, we control a few more variables related to literacy rates (m3). The effect of open defecation drops further but remains significant at 5 percent level. However, instead of MPCE, overall literacy rates, women literacy rates and household sizes dominate as significant factors affecting stunting. This specification is comparable with model 4 of \citep{Spears2013}, but the estimates of the partial effect of open defecation remain very high in \citep{Spears2013}. A 10\% decrease in open defecation rate is associated with only 0.13 percentage point decrease in stunting rate whereas \citep{Spears2013} conclude that for their sample, the reduction is about 0.7 percent. We should keep in mind that the sample districts in \citep{Spears2013} are non-random and are selected based on high prevalence of stunting.

There are several other important covariates which we believe can have effect on stunting rates. In our final specification (m4) we  also include proportion of women having secondary or higher education, proportion of mothers who took folic acid supplement during pregnancy, proportion of mothers with full antenatal care, proportion of mothers who gave caesarean birth, proportion of women with below normal BMI, proportion of children who had vitamin A, and proportion of households with clean fuel source for cooking. As infant mortality rate is a problematic explanatory variable, we exclude it from our final specification. With this full set of controls, our estimation shows that literacy rates and education levels, particularly women having secondary or higher level of  education, are important factors negatively associated with stunting rate. Open defecation remains a significant variable but its effect is much lower as compared to that in \citep{Spears2013}. The main contrast between our results and that of \citep{Spears2013} is that for our sample education plays an important role in improving stunting, particularly through mothers' education, whereas in the  \citep{Spears2013} sample, education does not play a significant role.  Thus our findings are also  in  agreement with those in \citep{som-pal-bharti,kumar-sinha}. 

In Table ~\ref{Table:predictedvsactual} we compare the regression results using predicted and actual variables. They are qualitatively similar,  but the predicted variables are more ``noise free'' and provide more consistent and interpretable estimates.

\begin{figure}[h!]
\begin{center}
\includegraphics{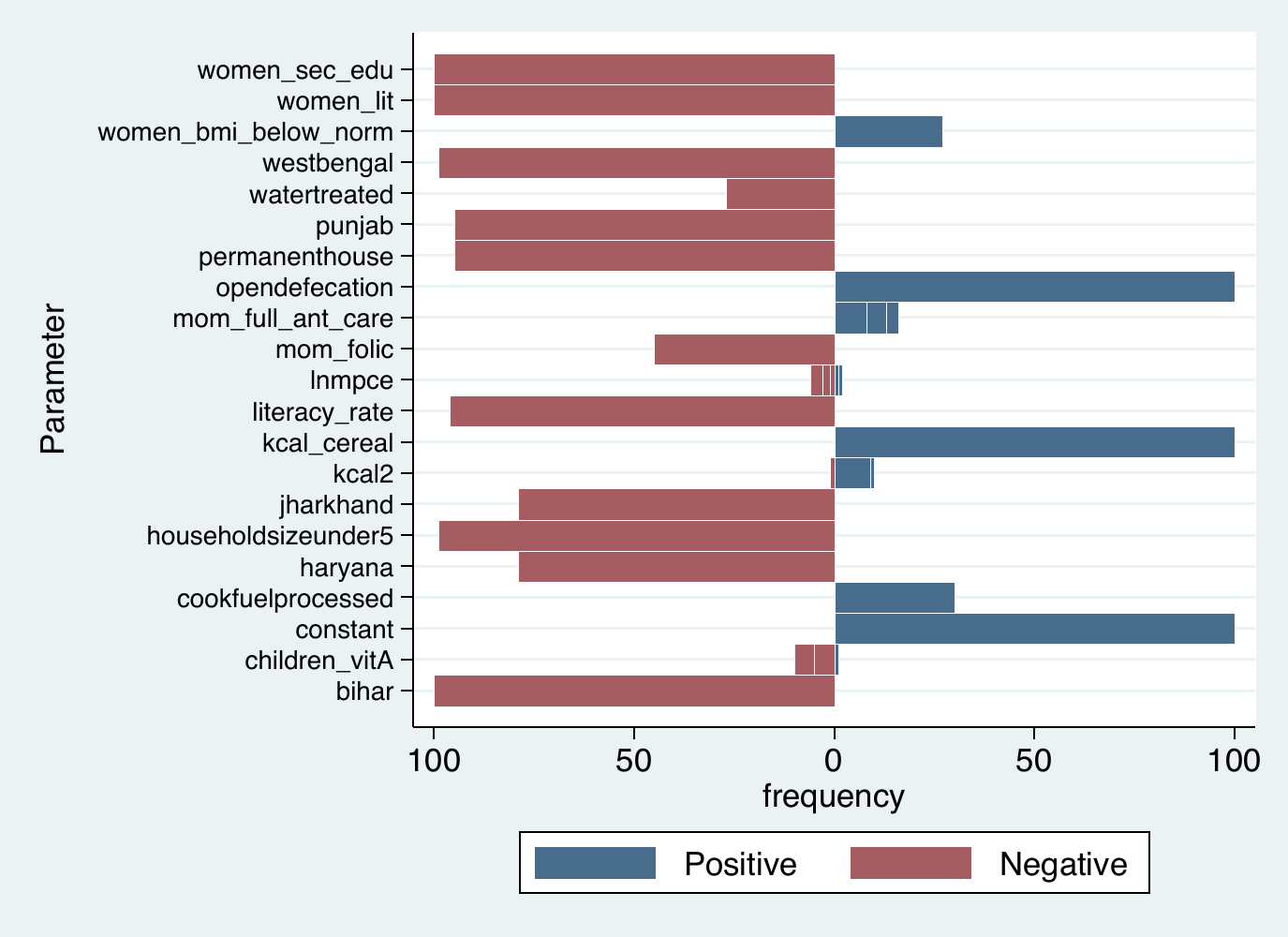}
\end{center}
\caption{Histogram of  the fraction of times a control variable is 10\%, 5\% an 1\% significant. The positive and the negative associations are shown separately. The average $R^2$ score for regression was 0.65.}
\label{fig:starplot}
\end{figure}

\begin{figure}[h!]
\begin{center}
\begin{tabular}{cc}
\includegraphics[width=2.5in]{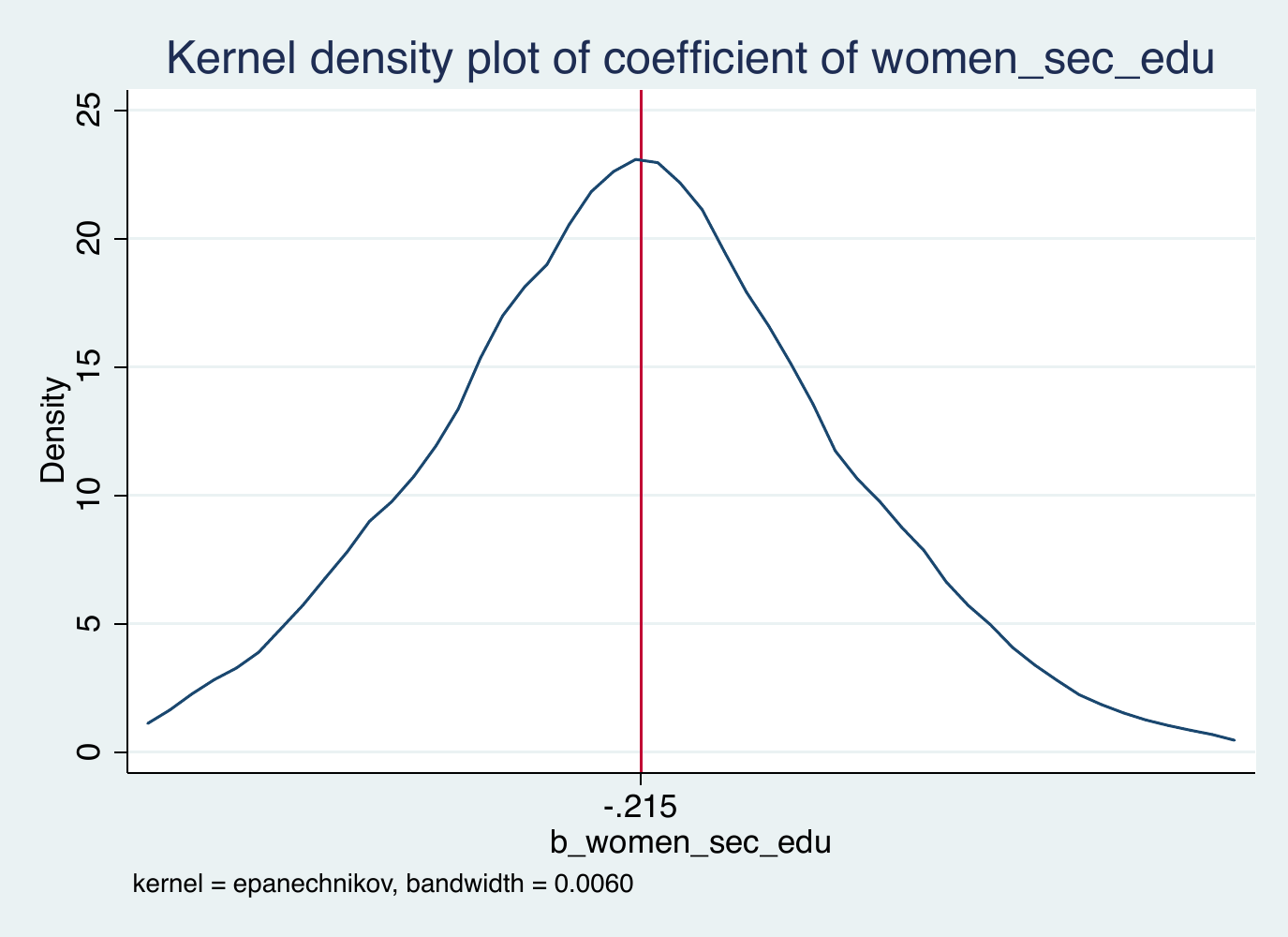} & \includegraphics[width=2.5in]{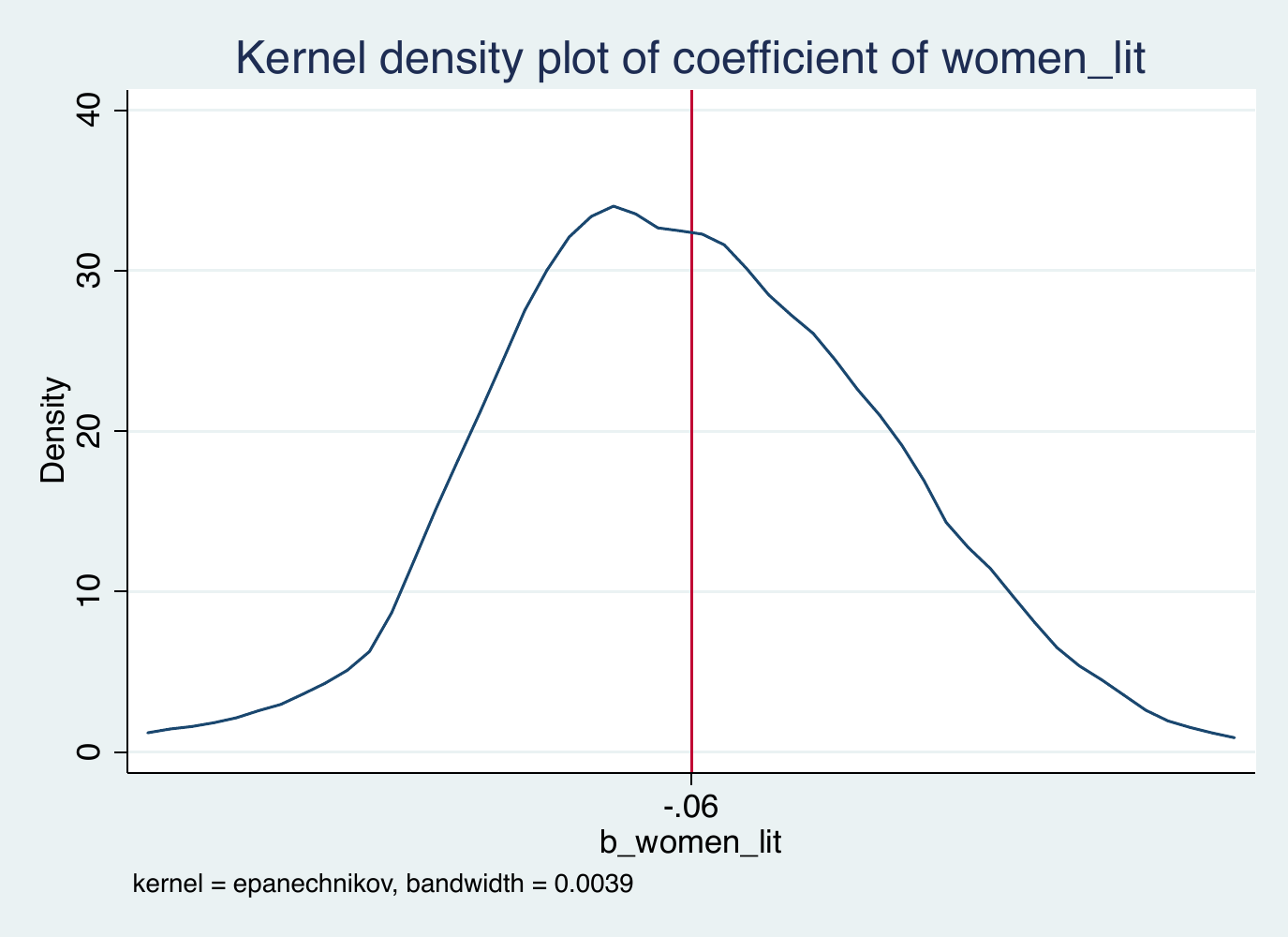}  \\
\includegraphics[width=2.5in]{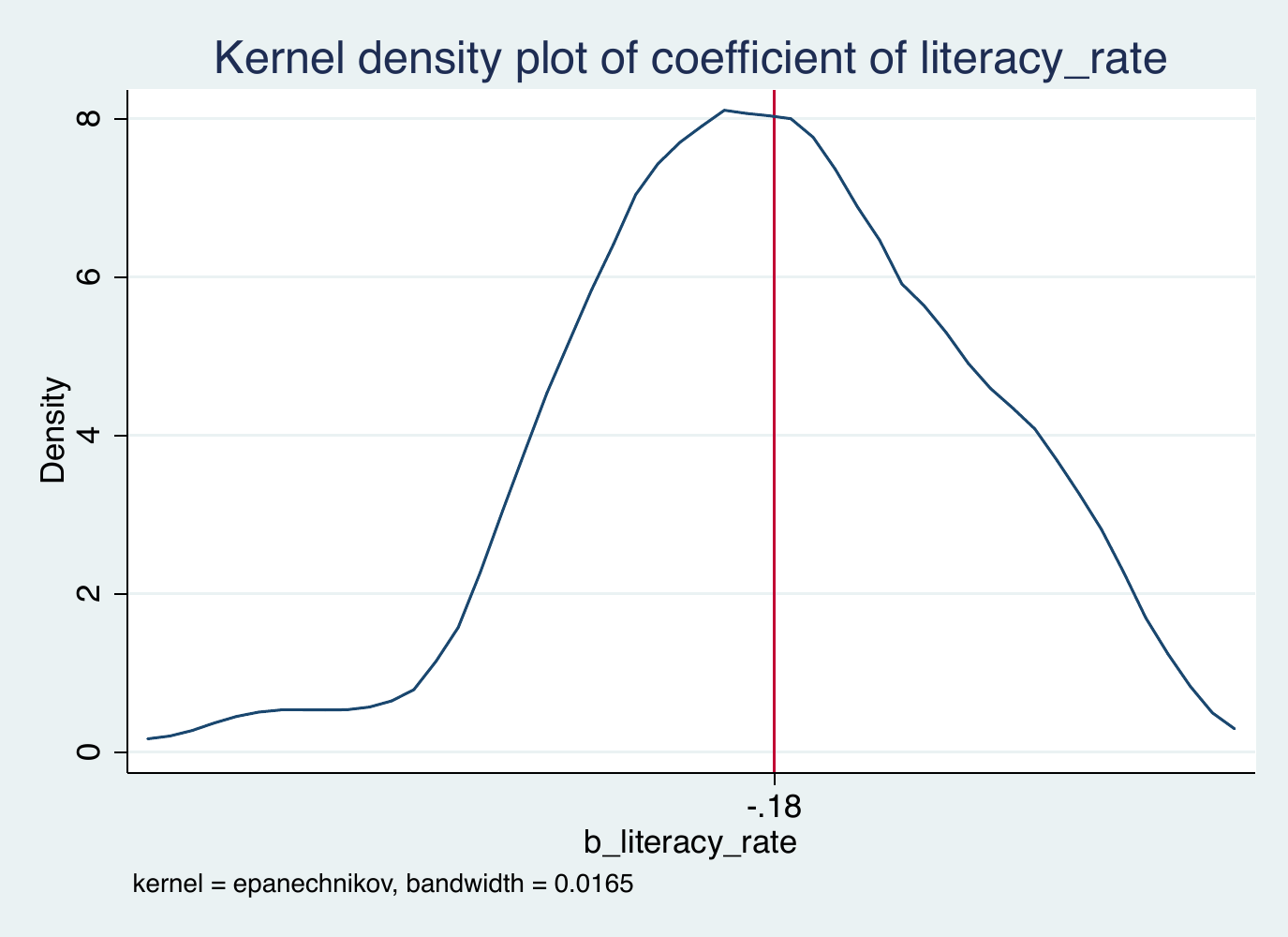} & \includegraphics[width=2.5in]{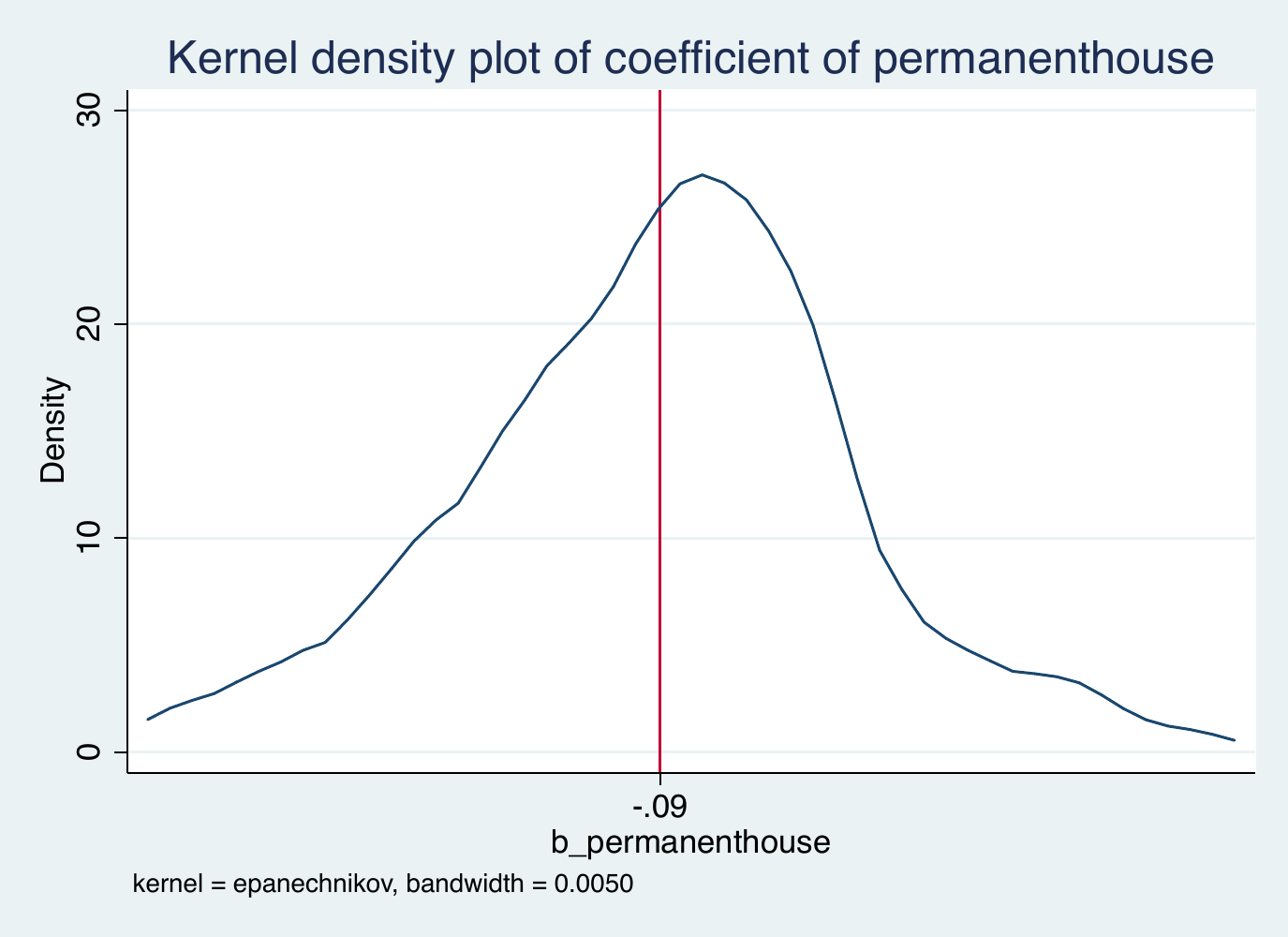}  \\
\includegraphics[width=2.5in]{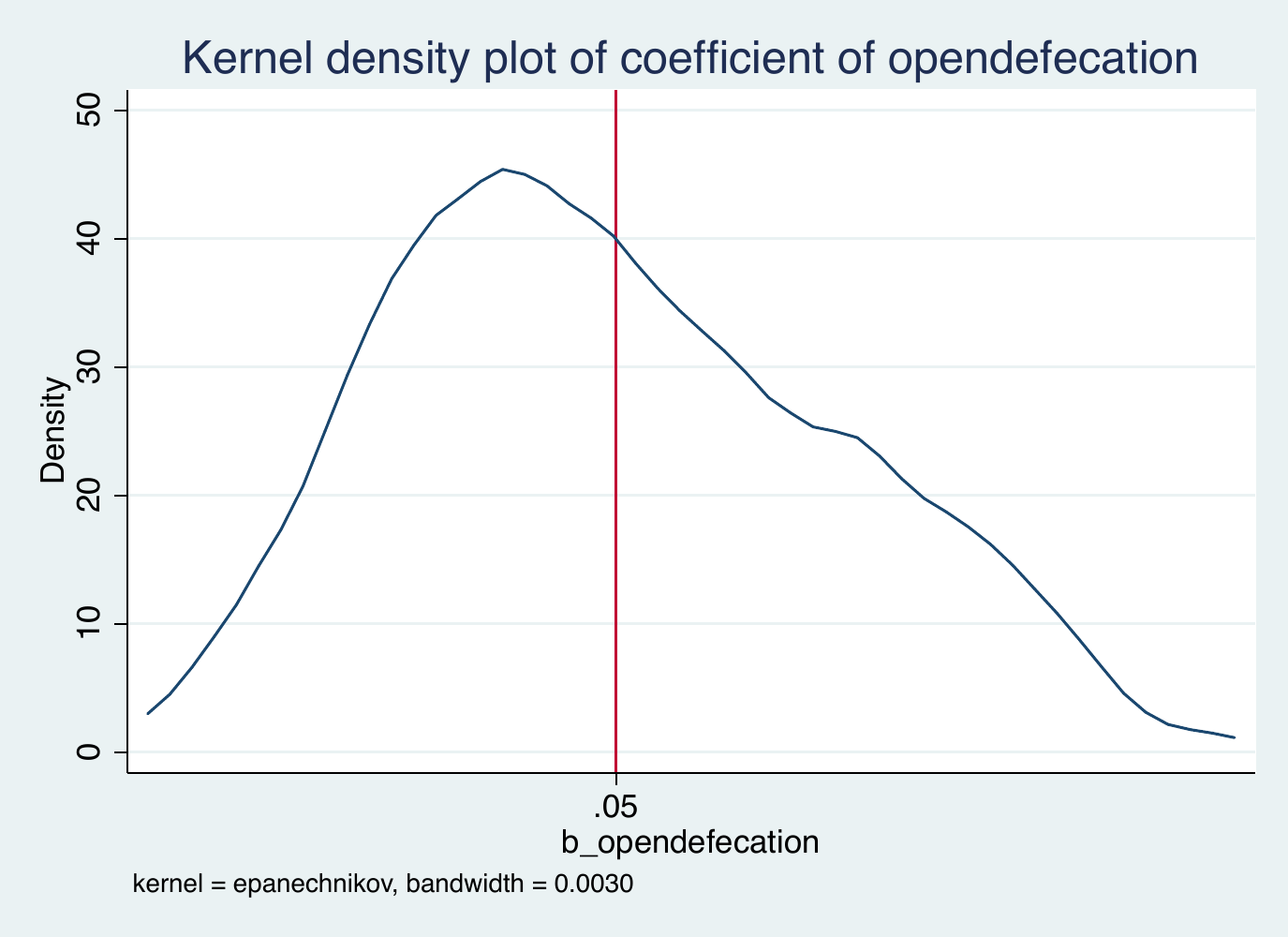} & \includegraphics[width=2.5in]{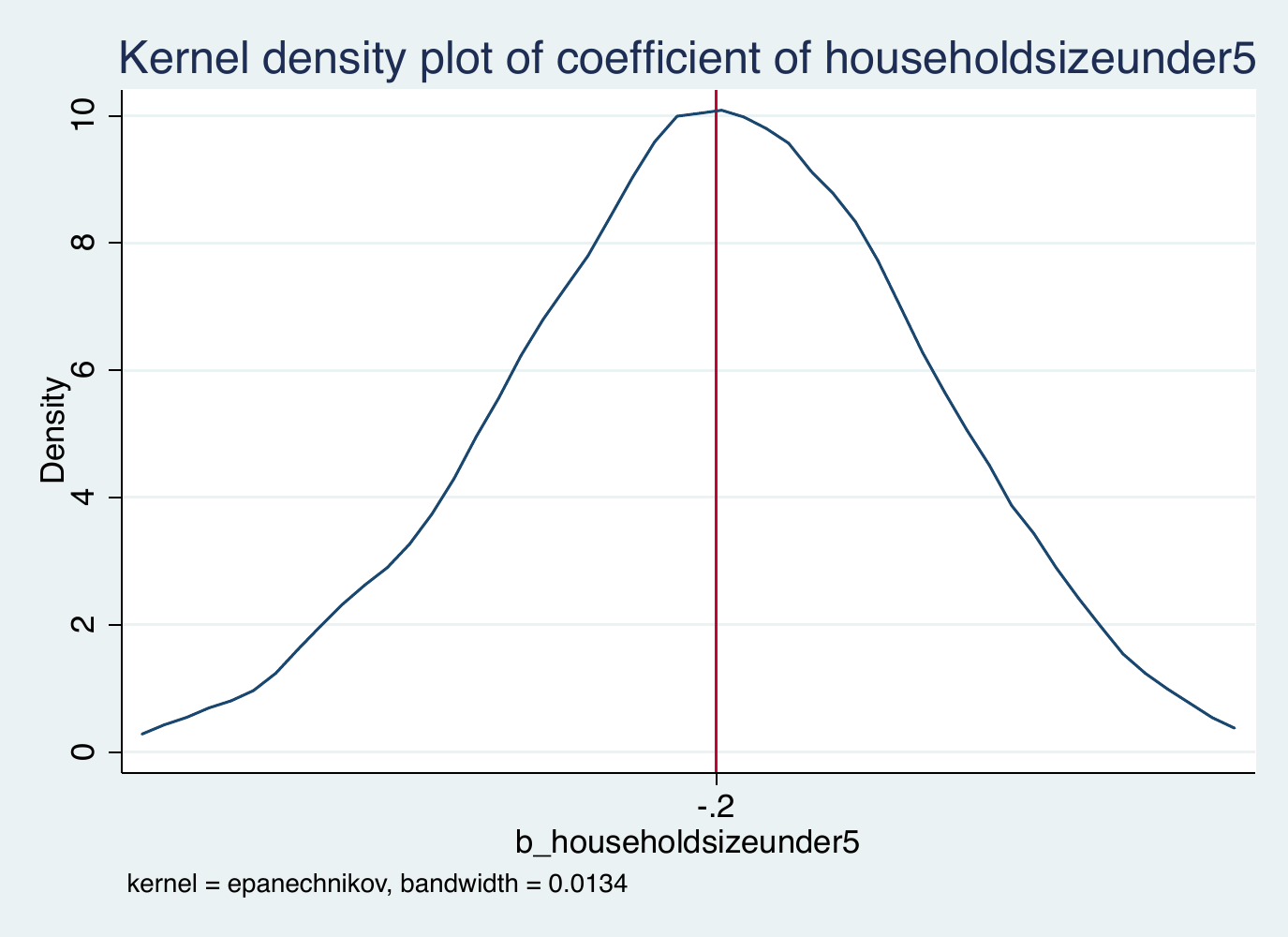}  \\
\includegraphics[width=2.5in]{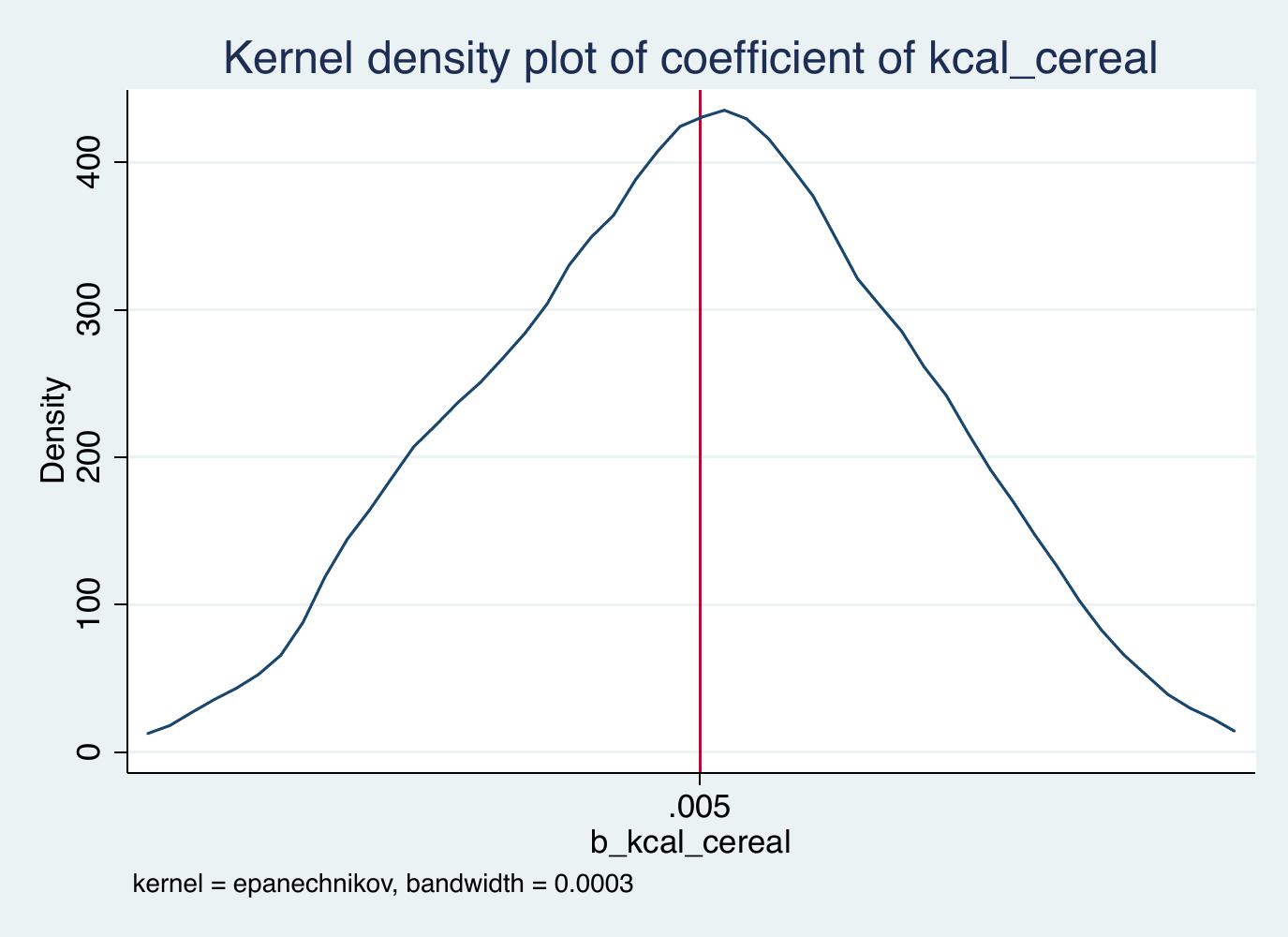} & \includegraphics[width=2.5in]{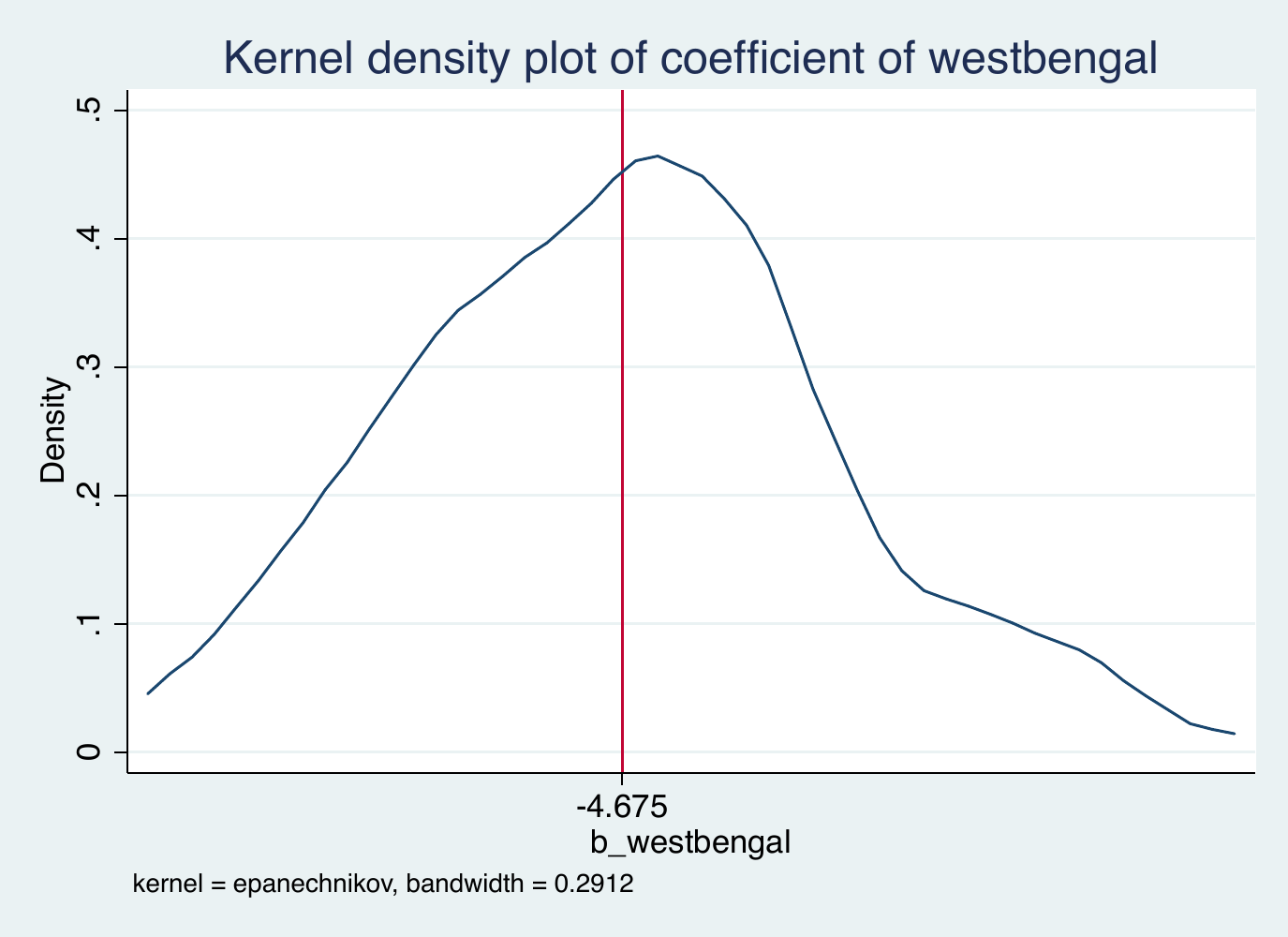}  
\end{tabular}
\caption{Kernel density plots of regression coefficients for some of the significant explanatory variable. We show the density plot for only one state (categorical variable), the others are similar.}
\label{fig:kdensity}
\end{center}
\end{figure}

Note that all the above regression analyses, either by \citep{Spears2013}  or by us, were carried out using  district level aggregated data. Most of the survey variables, including stunting rates, are available only at the district level. However, we can use our machine learning tool to predict corresponding village level estimates from satellite images. Consequently, we can also carry out the same regression analysis at a much lower level of data aggregation at the village level. Our six north Indian states have 2,18,000 villages, and linear regression with a such a large sample size makes the $p$ values go to zero very quickly. In view of this we take uniform random samples of villages multiple times and do repeated regression analysis. We carry out a power and sample size estimation using the procedure outlined in \citep{faul2007g}, and determine the optimal sample size for regression to be approximately 3500 villages. We carry out the regression analysis 100 times using 3500 randomly chosen villages and compute the histogram of 1\%, 5\% and 10\% significance levels for each variable and the kernel densities of the corresponding regression coefficients. For the village level analysis we introduce the proportion of permanent houses in a village as an additional control which measure a village level asset infrastructure. For the village level regression we also control for the states keeping the largest state, Uttar Pradesh, as the base (note that UP also has the highest rate of stunting).  In Figure ~\ref{fig:starplot} we plot the histogram of  the fraction of times a variable is 10\%, 5\% an 1\% significant. We show the positive and the negative associations separately. In Figure ~\ref{fig:kdensity} we show the kernel density plots of the regression coefficients for the significant variables. We show it for only one state, West Bengal. 

The village level analysis clearly reinforces that although the rate of open defecation is an important factor, women's education levels are the most significant determinant of stunting. Moreover, there are  several other infrastructure and diet related explanatory variables that turn out to be at least as important as open defecation. The limitation of the sample in \citep{Spears2013} due to unavailability of data for the whole set of districts (or a representative sample of districts), led them to quite a different conclusion. In contrast, using satellite based predictions we generate quite accurate measures of some more controls and construct a representative sample of districts. Using this richer sample, our conclusion is that along with open defecation, women education is also an important factor that affects stunting. We also confirm this by detailed village level analysis using predicted data.

\section{Conclusion}
\label{sec:conclu}
We have presented a tool for monitoring development indicators using high resolution day-time satellite images which can provide valuable estimates when survey data are infrequent or missing. We use machine learning to build a deep CNN based regression model for a hand crafted asset vector from input satellite images. Though the model is static and is trained with cross-sectional data, we demonstrate that it can be effectively used to predict the asset model from satellite images acquired at different times, making it an extremely useful  alternative between surveys. Further, the asset model can be used for transfer learning and prediction of a variety of other socio-economic and health parameters, and we demonstrate the use of predicted variables using a regression case study to understand the determinants of stunting. We also demonstrate an interesting application of our tool to generate alerts by detecting dissimilar regression outcomes in geographically neighbouring regions.  Predictive tools such as the one we have presented may find wide use in validation of surveys, in monitoring development progress of regions and in evidence based design of social policies.

\section*{Acknowledgement}
We thank Armaan Bhullar for research assistance. Part of this work was initiated during his Bachelor's project at IIT Delhi.

We thank the IIT Delhi High Performance Computing facility for providing the computational resources for this work.

We thank Dr. Manish Sharma and M/s Pitney Bowes (\href{https://www.pitneybowes.com/in}{https://www.pitneybowes.com/in}) for providing us with the proprietary shape files for all the villages in the six north Indian states, indexed by the village census ids.
\bibliographystyle{plainnat}

\bibliography{satellite}

\newpage
\appendix
\section{Appendix: Data description}
\label{sec:data}
In what follows we briefly describe the data that we have used for our computational experiments.

\subsection{Asset model}
\label{sec:asset}

\begin{table}[ht]
\centering
\caption{Asset indicators aggregated from village level Census 2011 data}
\label{table:assets}
{\tiny
\begin{tabular}{|l|l|l|}
\hline
Indicator & Description (\% of houses in the village with) & Aggregated from columns \\
\hline \hline
electronics & radio/transistor/tv/laptop & ([128]+[129]+[130]+[131)]/3\\ \hline
water-treated & water from treated source/covered well/ tube-well & [72]+[74]+[77] \\  \hline
water-untreated & water from untreated source/uncovered well & [73]+[75]\\  \hline
water-natural & drinking water from ponds/rivers/lakes & [76]+[78]+[79]+[80]+[81] \\  \hline
light from electricity &  from grid/solar & [85]+[87] \\  \hline
light from oil & kerosene/other oil & [86]+[88]+[89] \\  \hline
has-phone &  land-line/mobile/both & [132]+[133]+[134] \\  \hline
transport-cycle &  cycle & [135] \\  \hline
transport-motorized &  motorcycle/scooter/car/jeep & [136]+[137] \\  \hline
no-assets & no assets (cycle/phone etc.) & [139] \\  \hline
banking-services & availing banking services & [127] \\  \hline
cook-fuel-processed &  LPG/electric stove etc. & [113]+[114]+[115]\\   \hline
bathroom-within & bathroom within premises & [103]+[104] \\  \hline
rooms-under-3 &  less than 3 rooms & [49]+[50]+[51] \\  \hline
household-size-under-5 & less than 5 family members & [56]+[57]+[58]+[59] \\  \hline
permanent-house & permanent house & [140] \\  \hline
\end{tabular}
}
\end{table}
We create our asset model from the {\it Houselisting and Housing} data (Table HH-14) of Census 2011 \citep{census} corresponding to the six north Indian states of Punjab, Haryana,  Uttar Pradesh, Bihar, Jharkhand and West Bengal. This village level data is indexed by the village ids and provides aggregated information of about 140 amenities and assets in the village households. Each column indicates the percentage of households with a facility. We do a dimensionality reduction by manually aggregating, using weighted sums,  from the 140 dimensional vector to create a 16 dimensional vector (Table \ref{table:assets}).  

\begin{figure}[h!]
	\centering
    \includegraphics[scale=0.45]{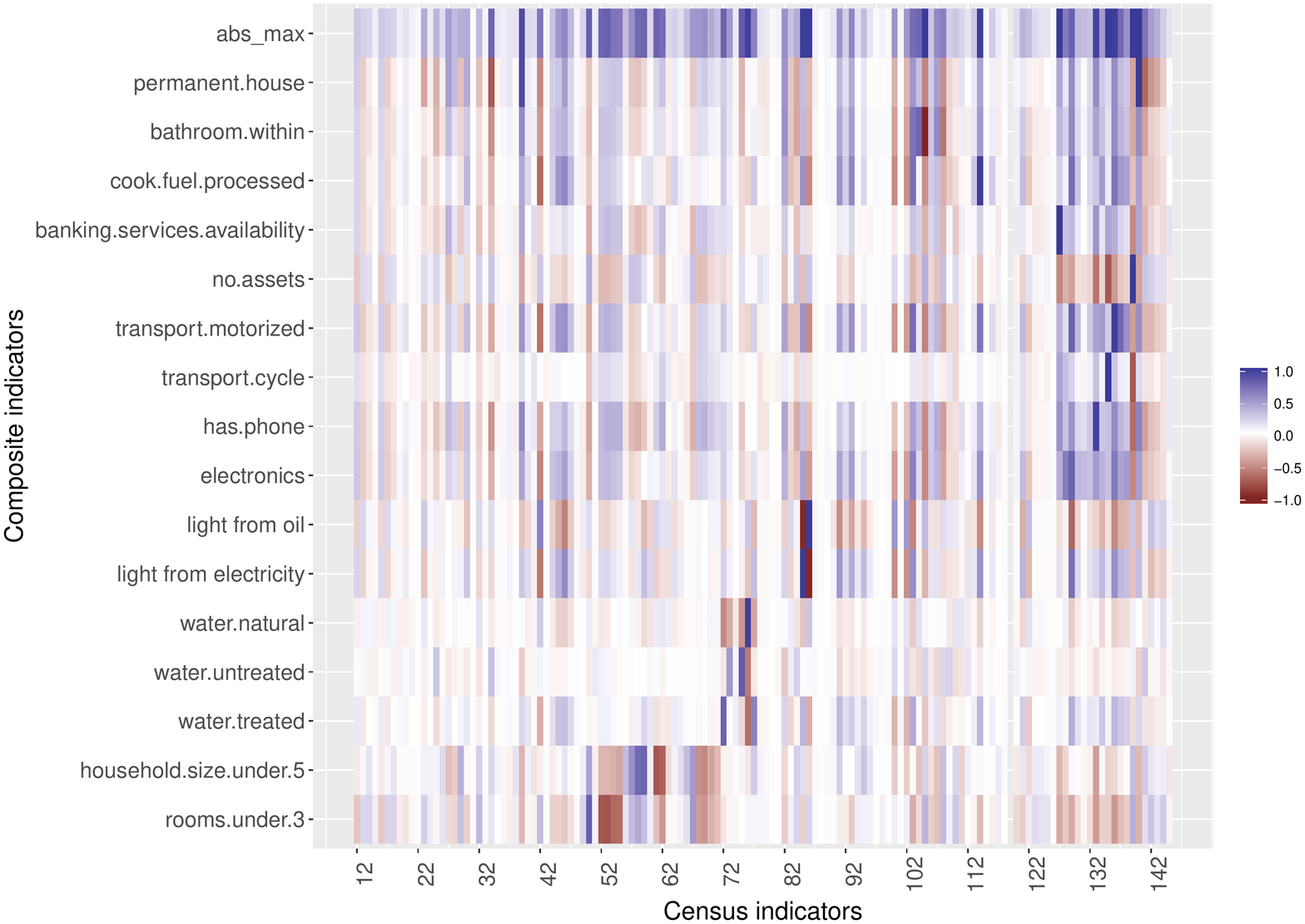}
    \caption{Correlation between each of the 140 census asset parameters and 16 aggregated indicators. Few census parameters have small numerical values and are hence poorly correlated with any of the 16 hand-crafted indicators.}
    \label{fig:assetcorrelation} 
\end{figure}

\citep{Jean790} used the first principal component of the data as the target of the regression.  However, we find that our manual aggregation gives better regression accuracy than using the linear principal component analysis (PCA) for dimensionality reduction.  Though the leading principal components capture high variance in the data, other components may be equally important for economic analysis.
In Figure \ref{fig:assetcorrelation}  we show the correlation matrix of the 140 column census asset data  with  the 16 aggregated indicators. Majority of the census parameters are strongly  correlated to one or more of the aggregated indicators. A few have small numerical values and are hence poorly correlated. This demonstrates that the 16 dimensional asset vector is representative of the original 140 dimensional asset data. 

The Census 2011 data is noisy and there are large errors for some villages; see Figure \ref{fig:regwb} for example. For rejecting outliers we compute the distribution of Mahalanobis distances of all villages. We find a steep rise above a distance threshold of 30 and we reject these villages as outliers. Approximately 5\% of the villages get rejected. In Section \ref{sec:regression} we present regression results with and without outlier rejection.

\subsection{Night lights}
We use the night light data provided by  the Defense Meteorological Satellite Program's Operational Linescan System (DMSP-OLS) \citep{night}. The night light data is available in 30 arc second grids, spanning $-180^\circ$ to $180^\circ$ longitude and $-65^\circ$ to $75^\circ$ latitude. Each 30 arc second grid cell is mapped to discrete values from $\{0,1,\ldots ,63\}$, where 63 corresponds to the highest night light intensity.  

\subsection{Daytime satellite images}
For geo-registration of the villages we use the polygon boundaries generated by Survey of India \citep{surveyofindia} \footnote{Actually, we use proprietary geo-registered boundary data obtained from M/s Pitney Bowes (\href{https://www.pitneybowes.com/in}{https://www.pitneybowes.com/in}), which is a post-processed and cleaned up version of the original Survey of India \citep{surveyofindia} data indexed by the village census ids. The village boundary data (shapefiles) are also publicly available at \href{https://github.com/justinelliotmeyers/official_india_2011_village_boundary_polygons_BHUVAN_2}{https://github.com/justinelliotmeyers/official\_india\_2011\_village\_boundary\_polygons\_BHUVAN\_2}.}. 
The village shape files  are linked to the  Census 2011 data  by the unique 16 digit census ids.

We obtain the daytime satellite images corresponding to the villages from Google static maps using the API provided by Google \citep{googleapi}. The Google static maps are free of cloud cover and other noise, because they are carefully constructed mosaics over time. The Google static maps only has recent imagery corresponding to 2017, which we use for regression of the census asset model corresponding to 2011. We assume that the visual characteristics have not changed significantly in six years to make the disparate cross-sectional mapping invalid. We comment more on this in Section \ref{sec:temporal}. We have experimented with both tiling the village shapes using 1 {\it Km}$^2$ images and using single images corresponding to the village centroid covering larger areas of  4 and 7 {\it Km}$^2$. Tiling the villages with 1{\it Km}$^2$ tiles is computationally expensive and gave similar regression scores compared  to single images at the village centroid.  Consequently, centred images were used in the final analysis.

\listofchanges
\listoftodos

\end{document}